\documentclass[a4paper]{article}

\usepackage{latexsym}
\usepackage[UKenglish]{babel}
\usepackage{graphicx}
\usepackage{algorithm}
\usepackage{algpseudocode}
\usepackage{epstopdf}
	\graphicspath{{./}{figures/}}
\usepackage[colorlinks=true,linkcolor=blue,citecolor=blue]{hyperref}
\usepackage{xcolor}
\usepackage{amsfonts,amsmath,amsthm,mathtools,bbm,cool}
\usepackage{tikz}
\usepackage{subcaption}
\usepackage{graphicx} 
\usepackage{array}
\usepackage{booktabs}
\usepackage{tabularx}

\newtheorem{Definition}{Definition}[section]
\newtheorem{Proposition}{Proposition}[section]

\theoremstyle{remark}

\newcommand{\be}{\begin{equation}}
\newcommand{\ee}{\end{equation}}

\allowdisplaybreaks

\begin{document}

\title{Understanding the Impact of Evaluation Metrics in Kinetic Models for Consensus-based Segmentation}

\author{R.F.Cabini\footnote{Euler Institute, Università della Svizzera Italiana; raffaella.fiamma.cabini@usi.ch} , H. Tettamanti\footnote{Department of Mathematics "F. Casorati", University of Pavia; horacio.tettamanti01@universitadipavia.it} , M. Zanella\footnote{Department of Mathematics "F. Casorati", University of Pavia; mattia.zanella@unipv.it}}  
\date{}
\maketitle

\begin{abstract}
In this article we extend a recently introduced kinetic model for consen\-sus-based segmentation of images. In particular, we will interpret the set of pixels of a 2D image as an interacting particle system which evolves in time in view of a consensus-type process obtained by interactions between pixels and external noise. Thanks to a kinetic formulation of the introduced model we derive the large time solution of the model. We will show that the choice of parameters defining the segmentation task can be chosen from a plurality of loss functions characterising the evaluation metrics.
\end{abstract}

\section{Introduction}

The primary objective of image segmentation is to partition an image into distinct pixel regions that exhibit homogeneous characteristics, including spatial proximity, intensity values, color variations, texture patterns, brightness levels, and contrast differences, thereby enabling more effective analysis and interpretation of the visual data. The application of image segmentation methods plays an important role in clinical research by facilitating the study of anatomical structures, highlighting regions of interest, and measuring tissue volume \cite{A_etal,BAJT,CMFPRC,M_etal,N_etal,SA}. In this context, the accurate recognition of areas  affected by pathologies can have great impact for more precise early diagnosis and monitoring in a great variety of disease that range from brain tumors to skin lesions. 

Over the past decades, a variety of computational strategies and mathematical approaches have been developed to address image segmentation challenges. Among these, deep learning techniques and neural networks have emerged as one of the most widely used methods in contemporary image segmentation tasks \cite{HJHK,IJKPMH,LSLZ,KWKP,RFB,YAZYT,ZRMNJ, LABC, LPB}. Leveraging a set of examples, these techniques are capable of approximating the complex nonlinear relationship between inputs and desired outputs. While deep learning models excel in complex segmentation problems, their dependence on large annotated datasets remains a significant challenge, particularly in fields such as biomedical imaging, where data availability is limited and manual labeling can be both expensive and time-consuming. A different approach is based on clustering methods \cite{CDA,FK,JMF,QFB,PGK,K}. These methods group pixels with similar characteristics, effectively partitioning the image into distinct regions. Clustering-based methods offer an attractive alternative to deep learning techniques, as they do not require supervised training and therefore, can be used on small, unlabeled datasets. In this direction , a kinetic approach for unsupervised clustering problems for image segmentation has been introduced in \cite{CPLFZ,HPV}. In these works, microscopic consensus-type models have been connected to image segmentation tasks by considering the pixels of an image as a interacting system where each particle is characterised by its space position and a feature determining the gray level. A virtual interaction between particles will then determine the asymptotic formation a finite number of clusters. Hence, a segmentation mask is generated by assigning the mean of their gray levels to each cluster of particles and by applying a binary threshold. Among the various nonlinear compromise terms that have been proposed in the literature, we will consider the Hegselmann-Krause model described in \cite{HK,RK2002} where it is supposed that each agent may only interact with other agent that are sufficiently close. This type of interaction is classically known as bounded confidence interaction function. As a result, two pixel will interact based on their distance in space and their gray level. The approach developed in \cite{CPLFZ} is based on the methods of kinetic theory for consensus formation. In the last decades, after the first model developed in \cite{D,DNAW,F,SWS}, several approaches have been designed to investigate the emergence of patterns and collective structures for large systems of agents/particles \cite{BT,CFL,MT14,FF}. To this end, the flexibility of kinetic-type equations have been of paramount importance to link the microscopic scale and the macroscopic observable scale \cite{APTZ,CFRT2010,DW2015,FR,PTTZ,PT2013,T}. 

In order to construct a data-oriented pipeline, we calibrate the resulting model by exploiting a family of existing evaluation metrics to obtain the relevant information from a ground truth image \cite{Auricchio,C,DI,J,MPSKPM,TH}. The main development of this study, compared to the one described in~\cite{BEBVMBB}, relies in the fact that we evaluate multiple metrics to quantify segmentation error, which is crucial for the optimization of the internal model parameters. In particular, we will concentrate on the  Standard Volumetric Dice Similarity Coefficient (Volumetric Dice), a volumetric measure based on the quotient between the intersection of the obtained segmented images and their total volume, the Surface Dice Similarity Coefficient, which is analogous to the Volumetric Dice but exploits the surface of the segmented images \cite{BEBVMBB}. Furthermore, we test the Jaccard index, which is an alternative option to evaluate the volumetric similarity between two segmentation masks, and the $F_\beta$-measure, which is a performance metrics which allows a balance between precision and sensitivity. 
In this paper we describe these metrics in detail and we analyze how these choices of evaluation metric influence the parameter optimization process. Furthermore, we discuss the most suitable metrics for the final assessment of the produced segmentations. This expanded evaluation provides novel insights into the impact of evaluation metrics on model performance and enhances our understanding of how to efficiently  optimize the introduced segmentation pipeline.

In more detail, the manuscript is organized as follows: In Section \ref{sect:HK} we introduce an extension of the Hegselmann-Krause model in 2D and we present the structure of the emerging steady states for different values of the model parameters. Next, we present a description of the model based on a kinetic-type approach. Furthermore, we show how can this model be extended and apply for the image segmentation problem. In Section \ref{sect:metrics} we present a Direct Simulation Monte Carlo (DSMC) Method to approximate the evolution of the system and we introduce possible optimization methods to produce segmentation masks for given images. To this end, we introduce the definition of the principal optimization metrics used in the context of bio-medical images and their principal characteristics. In Section \ref{sect:results} we show the results for a simple case of segmenting a geometrical image with a blurry background and compare the results obtained for different choices of the diffusion function. 
Finally, we present the results obtained for various Brain Tumor Images and discuss how the choice of different metrics may affect the final result. We show that the $F_{\beta}-$measure does not produce consistent results for different values of $\beta$. We reproduce the expected relationship between the Volumetric Dice Coefficient and Jaccard Index and show that both metrics plus the Surface Dice Coefficient yield similar results. Nevertheless, we argue that for this type of images the Surface Dice Coefficient produces more accurate loss values and its definition is more representative compare to the Volumetric Dice Coefficient and Jaccard Index.

\section{Consensus modelling and applications to image segmentation}\label{sect:HK}

In recent years, there has been growing interest in exploring consensus formation within opinion models to gain a deeper understanding of how social forces affect nonlinear aggregation processes in multiagent systems. To this end, various models have been proposed considering different scenarios and hypothesis on how the pairwise interactions may lead to the emergence of a position. For a finite number of particles, the dynamics is usually defined in terms of first order differential equations having the general form 
\begin{equation}
\label{eq:model_general}
\dfrac{d \textbf{x}_i}{dt} = \dfrac{1}{N} \sum_{j=1}^N P(\textbf{x}_i,\textbf{x}_j)(\textbf{x}_j-\textbf{x}_i), 
\end{equation}
where $\textbf{x}_i(t) \in \mathbb R^{d}$, $d\ge1$,  characterise the position of the agent $i= 1,\dots,N$ at time $t\ge0$, and $P(\cdot,\cdot)\ge0$ tunes the interaction between the agents $\textbf{x}_i,\textbf{x}_j \in \mathbb R^{d}$, see e.g. \cite{APZ,CFL,HK,MT14,NGW}.

In addition to microscopic agent-based models, in the limit of an infinite number of agents, it is possible to derive the evolution of distribution functions characterising the collective behavior of interacting systems. These approaches, typically grounded in kinetic-type partial differential equations (PDEs), are capable of bridging the gap between microscopic forces and the emerging properties of the system, see \cite{PT2013}. 

\subsection{The 2D bounded confidence model}

We now consider the bidimensional case, $d = 2$, and we specify the interaction function based on the so-called bounded confidence model. In more detail, we consider $N \geq 2$ agents and define their opinion variable through a vector $\mathbf{x} = (x_{i}(t),y_{i}(t)) \in \mathbf{R}^{2}$, characterised by  initial states $\{\mathbf{x}_{1}(0),\dots,\mathbf{x}_{N}(0) \}$. Agents will modify their opinion as a result of the interaction with other agents only if  $|\mathbf{x}_{i} - \mathbf{x}_{j}| \leq \Delta$ , where $\Delta \geq 0$ is a given confidence level. Hence, we can write \eqref{eq:model_general} as follows

\begin{equation}
        \frac{d}{dt}\mathbf{x}_{i} = \frac{1}{N}\sum_{j=1}^{N} P_\Delta (\mathbf{x}_i,\mathbf{x}_j)(\mathbf{x}_{j} - \mathbf{x}_{i}),
\end{equation}
where $P_\Delta (\mathbf{x}_i,\mathbf{x}_j) =\chi(|\mathbf{x}_{i} - \mathbf{x}_{j}| \leq \Delta): \mathbb R^2 \to \{0,1\}$, being $\chi(A)$ the characteristic function of the set $A \subseteq \mathbb R^2$. We can easily observe that the mean position of the ensemble of agents is conserved in time, indeed
\begin{equation}
\label{eq:conservation_micro}
\dfrac{d}{dt}\sum_{i=1}^N \textbf{x}_i =  \dfrac{1}{N} \sum_{i,j=1}^N \chi(|\textbf{x}_i-\textbf{x}_j|\le \Delta) (\textbf{x}_j-\textbf{x}_i) = 0, 
\end{equation}
thanks to the symmetry of the considered bounded-confidence interaction function. The bounded confidence model converges to a steady configuration, meaning that the systems reaches consensus in finite time. The structure of the steady state depends on the value of $\Delta$, see \cite{PTTZ}. 

\begin{figure}
    \centering
    \includegraphics[width=1.05\linewidth]{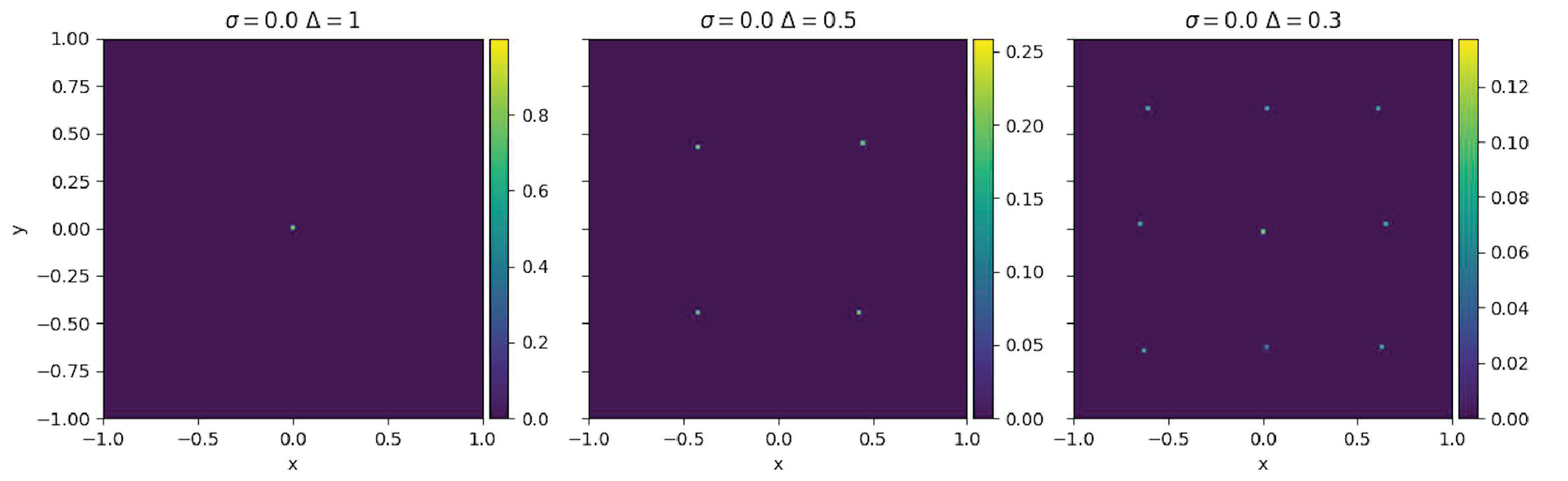} 
    \caption*{\centering (a)}
    \vspace{0.3cm}
    \centering
    \includegraphics[width=1.05\linewidth]{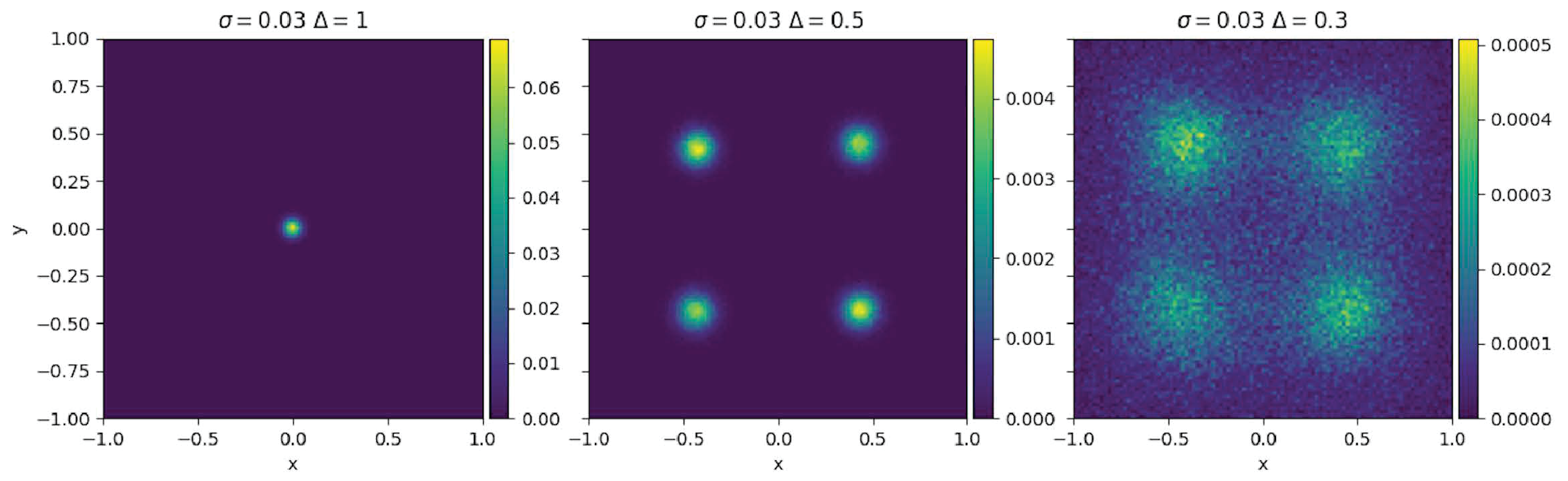} 
    \caption*{\centering (b)}
    \vspace{0.3cm}
    \centering
    \includegraphics[width=1.05\linewidth]{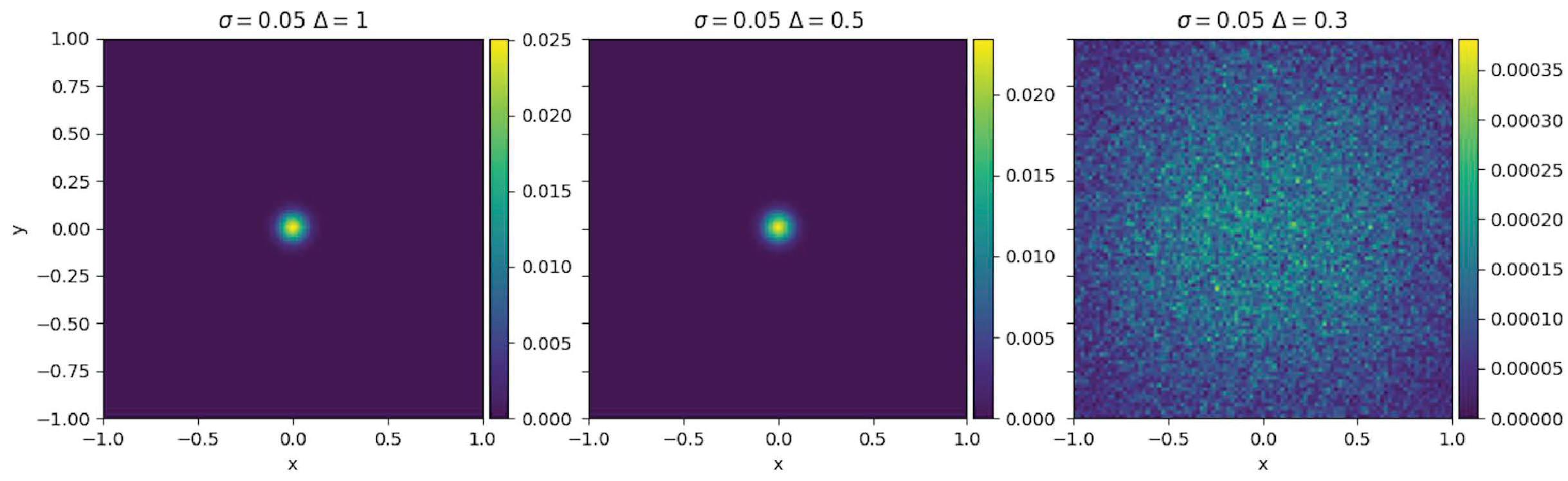} 
    \caption*{\centering (c)}
    \caption{ Large time distribution of the 2D bounded confidence model for different parameters characterising the compromise propensity and the diffusion for N = $10^5$ particles in $[0,T]$ with $T=100$ and $\Delta t = 0.01$. In (a) the final state converges to a number of clusters depending on the value of $\Delta$. As we reduce the range of interaction more clusters are created. In rows (b) and (c) we can see the interplay between the tendency of particles to aggregate and the diffusion. On the first column we see that the steady state converges to a Gaussian Distribution with a standard deviation given by $\sigma^{2}$. In the second column for (b) and (c) we see that the final states differ greatly in their structure. Finally, the last column shows the final states in the case where the diffusion surpasses considerable the aggregation tendency.  }
    \label{HK2D}
\end{figure}
Furthermore, to account for random fluctuations given by external factors in the opinion of agents we may consider a diffusion component as follows
\begin{equation} \label{timecontinousHK}
        d\mathbf{x}_{i} = \frac{1}{N}\sum_{j=1}^{N} P_{\Delta}(|\mathbf{x}_{i} - \mathbf{x}_{j}| \leq \Delta)(\mathbf{x}_{j} - \mathbf{x}_{i})dt + \sqrt{2 \sigma^{2}} d\mathbf{W}_{i}
\end{equation}
where $\{\mathbf{W}_{i}\}_{i=1}^N$ is a set of independent Wiener processes. The impact of the diffusion is weighted by the variable $\sigma^2>0$. To visualise the interplay between consensus forces and diffusion we depict in Figure~\ref{HK2D} the steady configuration of the model \eqref{timecontinousHK} for different combinations of the model parameters. For $\sigma^2 = 0$, the system forms a finite number of clusters depending on the value of $\Delta>0$, as illustrated in Figure~\ref{HK2D}(a). For values of the diffusion coefficient $\sigma^2>0$, the  number of clusters of the system varies as depicted in Figure~\ref{HK2D}(b). The right panel of Figure~\ref{HK2D}(b) shows the scenario in which the diffusion effect becomes comparable to the tendency of agents to cluster. Finally, in  Figure~\ref{HK2D}(c), for $\sigma = 0.05$, the diffusion effect dominates the grouping tendency, resulting in a homogeneous steady-state distribution.

\subsection{Kinetic models for consensus dynamics}\label{sect:consensus_kin}

In the limit $N\to +\infty$ it can be shown that the empirical density $$f^{(N)}(\textbf{x},t)= \frac{1}{N}\sum_{i=1}^N \delta(\textbf{x}-\textbf{x}_i(t))$$ of the system of particles \eqref{timecontinousHK} converges to a continuous density $f(\textbf{x},t): \mathbb R^2 \times \mathbb R_+ \to \mathbb R_+$ solution to the following mean-field equation
\begin{equation} \label{FP1}
\begin{split}
    \partial_{t} f(\mathbf{x},t) &= \nabla_{\textbf{x}} \cdot [\Xi[f](\mathbf{x},t) + \sigma^{2} \nabla_{\textbf{x}} f]    \\
    f(\mathbf{x},0) &= f_{0}(\mathbf{x})
\end{split}
\end{equation}
where $\Xi[f](\mathbf{x},t)$ is defined as follows
\begin{equation}
    \Xi[f](\mathbf{x},t) = \int_{\mathbf{R}^{2}} P_{\Delta}(\mathbf{x}_i,\mathbf{x}_j)(\mathbf{x} - \mathbf{x}_{*}) f(\mathbf{x}_*,t) d\mathbf{x}_{*}, 
\end{equation}
see e.g. \cite{CFTV}. 

We can derive \eqref{FP1} using a kinetic approach by writing $\textbf{x} := \textbf{x}_i(t)$ and $\textbf{x}_* := \textbf{x}_j(t)$ for a generic pair $(i,j)$ of interacting agents/particles and we approximate the
time derivative in \eqref{timecontinousHK} in a time step $\epsilon = \Delta t>0$, through an Euler-Maryuama approach, in the same spirit as \cite{CFRT2010,PTZ}. Hence, we recover the binary interaction rule

\begin{equation} \label{bi.scheme}     
\begin{split} 
        \mathbf{x}^\prime& = \mathbf{x} + \epsilon P_{\Delta}(\mathbf{x},\mathbf{x_{*}})(\mathbf{x_{*}} - \mathbf{x}) + \sqrt{2 \sigma^{2}} \eta \\
        \mathbf{x_{*}'} &= \mathbf{x_{*}} + \epsilon P_{\Delta}(\mathbf{x_{*}},\mathbf{x})(\mathbf{x} - \mathbf{x_{*}}) + \sqrt{2 \sigma^{2} } \eta_*, 
\end{split} 
\end{equation}
where $\mathbf{x}^\prime = \mathbf{x}_i(t+\epsilon)$,  $\mathbf{x}_*^\prime = \mathbf{x}_j(t+\epsilon)$
and $\eta,\eta_*$ are two independent 2D centered Gaussian distribution random variable such that
\begin{equation}
\label{eq:variance_scaling}
        \langle\eta\rangle = \langle\eta_*\rangle = 0 \qquad \langle\eta^2\rangle = \langle\eta_{*}^2\rangle = \epsilon 
\end{equation}
where $\langle\cdot\rangle$ denotes the integration with respect to the distribution $\eta$. 
Furthermore, in \eqref{bi.scheme} we shall consider $P(\mathbf{x},\mathbf{x}_{*}) = \chi (|\mathbf{x} - \mathbf{x_{*}}| < \Delta)$. We can remark that, if  $\sigma = 0$, since $P_{\Delta} \in [0,1]$ and $\epsilon \in (0,1)$ we get

\begin{equation} \label{mean}
\begin{split}
        \left\langle \mathbf{x' + x'_{*}} \right\rangle &= \mathbf{x + x_{*}} + \Delta t ( P_{\Delta}(\mathbf{x,x_{*}}) - P_{\Delta}(\mathbf{x_{*},x}))(\mathbf{x_{*} - x}) = \\
    &= \mathbf{x + x_{*}}
\end{split}    
\end{equation}
since the interaction function $P_{\Delta}$ is symmetric, consistently with \eqref{eq:conservation_micro}. This shows that the mean position is conserved at every interaction. Finally, we  have
\begin{equation} \label{eq.diss}
\begin{split}
         \mathbf{|\langle x'\rangle|^{2} + |\langle x\rangle|^{2}}  = \mathbf{|x'|^{2} + |x|^{2}} - 2 \Delta t P_{\Delta} \mathbf{|x' - x|^{2}} + o(\Delta t) 
\end{split}
\end{equation}
and the mean energy is dissipated at each interaction, since $P_{\Delta} \geq 0$. Hence, we consider the  distribution function $f = f(\mathbf{x},t): R^{2} \times R_{+} \rightarrow R_{+}$, such that $f(\mathbf{x},t)d\mathbf x$ represents the fraction of agents/particles in $[{x}_1,{x}_1 + d{x}_1) \times [{x}_2,{x}_2 + d{x}_2]$ at time $t\ge0$. The evolution of $f$ as a result of binary-interaction scheme \eqref{bi.scheme} is obtained by a Boltzmann-type  equation which reads in weak form 
\begin{equation}\label{eq:boltzmann}
\begin{split}
    &\dfrac{d}{dt} \int_{\mathbb R^2}\varphi(\mathbf{x})f(\mathbf{x},t)d\mathbf{x}  = \\
    &\quad  \left \langle \int_{\mathbb R^{4}} (\varphi(\mathbf{x}^\prime) - \varphi(\mathbf{x})) f(\mathbf{x},t)f(\mathbf{x}_*,t)d\mathbf{x} d\mathbf{x}_{*} \right \rangle,
\end{split}
\end{equation}
    being $\varphi(\cdot)$ a test function. As observed in \cite{T}, when $\Delta t = \epsilon \to 0^+$ we can observe that the binary scheme  \eqref{bi.scheme} becomes quasi-invariant and we can introduce the following expansion
    \begin{equation}
    \label{eq:expa}
    \left\langle\varphi(\mathbf{x}^\prime) - \varphi(\mathbf{x}) \right\rangle = \langle \mathbf x^\prime - \mathbf x\rangle \cdot \nabla_\mathbf{x} \varphi(\mathbf x) + \dfrac{1}{2} \left\langle (\mathbf{x}^\prime - \mathbf{x})^T H[\varphi] (\mathbf{x}^\prime - \mathbf{x})\right\rangle + R_\epsilon(\mathbf{x},\mathbf{x}_*) 
    \end{equation}
    being $R_\epsilon(\mathbf{x},\mathbf{x}_*)$ a reminder term and $H[\varphi]$ the Hessian matrix. Hence, scaling $\tau = \epsilon t$ and the distribution $f_\epsilon(\mathbf{x},\tau) = f(\mathbf{x},\tau / \epsilon)$, we may plug \eqref{eq:expa} in \eqref{eq:boltzmann} to get
    \[\begin{split}
\dfrac{d}{d\tau} \int_{\mathbb R^2} \varphi(\mathbf{x})f_\epsilon(\mathbf{x},t)d\mathbf{x} =& \dfrac{1}{\epsilon} \int_{\mathbb R^4} \langle \mathbf{x}^\prime - \mathbf{x} \rangle \cdot \nabla_\mathbf{x}\varphi(\mathbf{x})f_\epsilon(\mathbf{x},\tau)f_\epsilon(\mathbf{x}_*,\tau)d\mathbf{x}d\mathbf{x}_*+ \\
& \dfrac{1}{2\epsilon}\int_{\mathbb R^4} \langle (\mathbf{x}^\prime-\mathbf{x})^T H[\varphi(\mathbf{x}](\mathbf{x}^\prime-\mathbf{x})\rangle f_\epsilon(\mathbf{x},\tau)f_\epsilon(\mathbf{x}_*,\tau)d\mathbf{x}d\mathbf{x}_* + \\
& \dfrac{1}{\epsilon} \int_{\mathbb R^4}R_\epsilon(\mathbf{x},\mathbf{x}_*)f_\epsilon(\mathbf{x},\tau)f_\epsilon(\mathbf{x}_*,\tau)d\mathbf{x}d\mathbf{x}_*
    \end{split}\]
    Following \cite{CPLFZ}, see also \cite{PT2013}, we can prove that 
    \[
    \int_{\mathbb R^4}R_\epsilon(\mathbf{x},\mathbf{x}_*)f_\epsilon(\mathbf{x},\tau)f_\epsilon(\mathbf{x}_*,\tau)d\mathbf{x}d\mathbf{x}_* \to 0^+,
    \]
    as $\epsilon\to 0^+$. Hence, integrating back by parts the first two terms we obtain \eqref{FP1}. In more detail, 
    we can prove that $f_\epsilon$ converges, up to extraction of a subsequence to a probability density $f(\mathbf x,\tau)$ that is weak solution to the nonlocal Fokker-Planck  equation \eqref{FP1}. 

\subsection{Application to Image Segmentation}

\begin{figure}
\centering
\begin{tikzpicture}

\definecolor{gray1}{rgb}{0.1,0.1,0.1}
\definecolor{gray2}{rgb}{0.3,0.3,0.3}
\definecolor{gray3}{rgb}{0.5,0.5,0.5}
\definecolor{gray4}{rgb}{0.7,0.7,0.7}
\definecolor{gray5}{rgb}{0.9,0.9,0.9}

\fill[gray1] (0,3) rectangle ++(1,-1);
\fill[gray2] (1,3) rectangle ++(1,-1);
\fill[gray3] (2,3) rectangle ++(1,-1);
\fill[gray4] (3,3) rectangle ++(1,-1);

\fill[gray2] (0,2) rectangle ++(1,-1);
\fill[gray4] (1,2) rectangle ++(1,-1);
\fill[gray5] (2,2) rectangle ++(1,-1);
\fill[gray3] (3,2) rectangle ++(1,-1);

\fill[gray3] (0,1) rectangle ++(1,-1);
\fill[gray1] (1,1) rectangle ++(1,-1);
\fill[gray4] (2,1) rectangle ++(1,-1);
\fill[gray2] (3,1) rectangle ++(1,-1);

\fill[gray5] (0,0) rectangle ++(1,-1);
\fill[gray2] (1,0) rectangle ++(1,-1);
\fill[gray3] (2,0) rectangle ++(1,-1);
\fill[gray1] (3,0) rectangle ++(1,-1);

\draw[black] (0, 3) rectangle (4, -1);

\draw[red, thick] (1,2) rectangle ++(1,-1);

\node[draw=blue!20, fill=blue!10, rounded corners, inner sep=5pt] at (5.8,1.5) {\Large $(x_i, y_i, c_i)$};

\draw[->, red, thick] (1.5,1.5) -- (4.5,1.5);

\draw[fill=gray1, thick] (8+0,3-0.5) circle (0.4);
\draw[fill=gray2, thick] (8+1,3-0.5) circle (0.4);
\draw[fill=gray3, thick] (8+2,3-0.5) circle (0.4);
\draw[fill=gray4, thick] (8+3,3-0.5) circle (0.4);

\draw[fill=gray2, thick] (8+0,3-1-0.5) circle (0.4);
\draw[fill=gray4, thick] (8+1,3-1-0.5) circle (0.4);
\draw[fill=gray5, thick] (8+2,3-1-0.5) circle (0.4);
\draw[fill=gray3, thick] (8+3,3-1-0.5) circle (0.4);

\draw[fill=gray3, thick] (8+0,3-2-0.5) circle (0.4);
\draw[fill=gray1, thick] (8+1,3-2-0.5) circle (0.4);
\draw[fill=gray4, thick] (8+2,3-2-0.5) circle (0.4);
\draw[fill=gray2, thick] (8+3,3-2-0.5) circle (0.4);

\draw[fill=gray5, thick] (8+0,3-3-0.5) circle (0.4);
\draw[fill=gray2, thick] (8+1,3-3-0.5) circle (0.4);
\draw[fill=gray3, thick] (8+2,3-3-0.5) circle (0.4);
\draw[fill=gray1, thick] (8+3,3-3-0.5) circle (0.4);

\draw[black] (0+8-0.5, 3) rectangle (4+8-0.5, -1);

\draw[red, thick] (8+1,3-1-0.5) circle (0.5);

\draw[->, red, thick] (7.1,1.5) -- (8+1,3-1-0.5);

\end{tikzpicture} 
\caption{ A schematic representation of the proposed model where each pixel can be interpreted as a particle $(x_{i},y_{i},c_i)$, being $c_i$ a static feature in the interval $[0,1]$. 
}
\label{ISHK}
\end{figure}
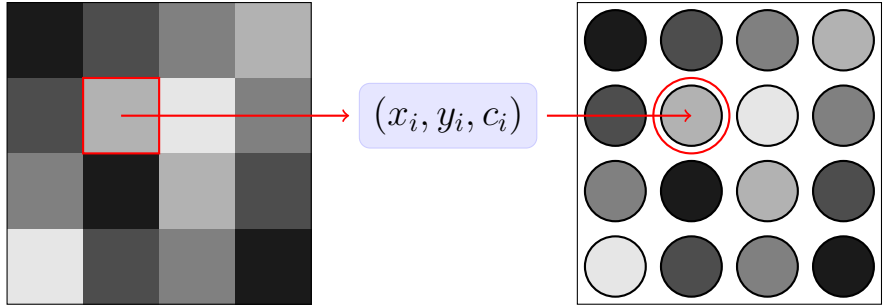

\begin{figure}[H]
\centering
\resizebox{\textwidth}{!}{\begin{tikzpicture}

\definecolor{gray1}{rgb}{0.1,0.1,0.1}
\definecolor{gray2}{rgb}{0.3,0.3,0.3}
\definecolor{gray3}{rgb}{0.5,0.5,0.5}
\definecolor{gray4}{rgb}{0.7,0.7,0.7}
\definecolor{gray5}{rgb}{0.9,0.9,0.9}

\draw[black] (-0.5, -1) rectangle (3.5, 3);
\draw[black] (0,3-0.5) circle (0.4); \fill[gray1] (0,3-0.5) circle (0.4);
\draw[black] (1,3-0.5) circle (0.4); \fill[gray2] (1,3-0.5) circle (0.4);
\draw[black] (2,3-0.5) circle (0.4); \fill[gray3] (2,3-0.5) circle (0.4);
\draw[black] (3,3-0.5) circle (0.4); \fill[gray4] (3,3-0.5) circle (0.4);

\draw[black] (0,3-1-0.5) circle (0.4); \fill[gray2] (0,3-1-0.5) circle (0.4);
\draw[black] (1,3-1-0.5) circle (0.4); \fill[gray4] (1,3-1-0.5) circle (0.4);
\draw[black] (2,3-1-0.5) circle (0.4); \fill[gray5] (2,3-1-0.5) circle (0.4);
\draw[black] (3,3-1-0.5) circle (0.4); \fill[gray3] (3,3-1-0.5) circle (0.4);

\draw[black] (0,3-2-0.5) circle (0.4); \fill[gray3] (0,3-2-0.5) circle (0.4);
\draw[black] (1,3-2-0.5) circle (0.4); \fill[gray1] (1,3-2-0.5) circle (0.4);
\draw[black] (2,3-2-0.5) circle (0.4); \fill[gray4] (2,3-2-0.5) circle (0.4);
\draw[black] (3,3-2-0.5) circle (0.4); \fill[gray2] (3,3-2-0.5) circle (0.4);

\draw[black] (0,3-3-0.5) circle (0.4); \fill[gray5] (0,3-3-0.5) circle (0.4);
\draw[black] (1,3-3-0.5) circle (0.4); \fill[gray2] (1,3-3-0.5) circle (0.4);
\draw[black] (2,3-3-0.5) circle (0.4); \fill[gray3] (2,3-3-0.5) circle (0.4);
\draw[black] (3,3-3-0.5) circle (0.4); \fill[gray1] (3,3-3-0.5) circle (0.4);

\draw[black] (4.5, -1) rectangle (8.5, 3);
\draw[black] (5+0.15,3-0.5-0.1) circle (0.4); \fill[gray1] (5+0.15,3-0.5-0.1) circle (0.4);
\draw[black] (6+0.1,3-0.5-0.2) circle (0.4); \fill[gray2] (6+0.1,3-0.5-0.2) circle (0.4);
\draw[black] (7-0.1,3-0.5-0.3) circle (0.4); \fill[gray3] (7-0.1,3-0.5-0.3) circle (0.4);
\draw[black] (8-0,3-0.5+0) circle (0.4); \fill[gray4] (8-0,3-0.5+0) circle (0.4);

\draw[black] (5+0.15,3-1-0.5+0.2) circle (0.4); \fill[gray2] (5+0.15,3-1-0.5+0.2) circle (0.4);
\draw[black] (6-0.2,3-1-0.5+0.3) circle (0.4); \fill[gray4] (6-0.2,3-1-0.5+0.3) circle (0.4);
\draw[black] (7-0.4,3-1-0.5+0.15) circle (0.4); \fill[gray5] (7-0.4,3-1-0.5+0.15) circle (0.4);
\draw[black] (8-0.1,3-1-0.5+0.2) circle (0.4); \fill[gray3] (8-0.1,3-1-0.5+0.2) circle (0.4);

\draw[black] (5,3-2-0.5) circle (0.4); \fill[gray3] (5,3-2-0.5) circle (0.4);
\draw[black] (6-0.3,3-2-0.5) circle (0.4); \fill[gray1] (6-0.3,3-2-0.5) circle (0.4);
\draw[black] (7+0.3,3-2-0.5) circle (0.4); \fill[gray4] (7+0.3,3-2-0.5) circle (0.4);
\draw[black] (8,3-2-0.5) circle (0.4); \fill[gray2] (8,3-2-0.5) circle (0.4);

\draw[black] (5,3-3-0.5) circle (0.4); \fill[gray5] (5,3-3-0.5) circle (0.4);
\draw[black] (6-0.1,3-3-0.5+0.2) circle (0.4); \fill[gray2] (6-0.1,3-3-0.5+0.2) circle (0.4);
\draw[black] (7+0.5,3-3-0.5+0.5) circle (0.4); \fill[gray3] (7+0.5,3-3-0.5+0.5) circle (0.4);
\draw[black] (8,3-3-0.5) circle (0.4); \fill[gray1] (8,3-3-0.5) circle (0.4);

\draw[black] (9.5, -1) rectangle (13.5, 3);
\draw[black] (10+0.2,3-0.5-0.2) circle (0.4); \fill[gray1] (10+0.2,3-0.5-0.2) circle (0.4);
\draw[black] (11+0.1,3-0.5-0.3) circle (0.4); \fill[gray2] (11+0.1,3-0.5-0.3) circle (0.4);
\draw[black] (12-0.5,3-0.5-0.4) circle (0.4); \fill[gray3] (12-0.5,3-0.5-0.4) circle (0.4);
\draw[black] (13-0.1,3-0.5-1.0) circle (0.4); \fill[gray4] (13-0.1,3-0.5-1.0) circle (0.4);

\draw[black] (10+0.3,3-1-0.5+0.4) circle (0.4); \fill[gray2] (10+0.3,3-1-0.5+0.4) circle (0.4);
\draw[black] (11-0.2,3-1-0.5+0.8) circle (0.4); \fill[gray4] (11-0.2,3-1-0.5+0.8) circle (0.4);
\draw[black] (12-0.8,3-1-0.5+0.5) circle (0.4); \fill[gray5] (12-0.8,3-1-0.5+0.5) circle (0.4);
\draw[black] (13-0.1,3-1-0.5-0.6) circle (0.4); \fill[gray3] (13-0.1,3-1-0.5-0.6) circle (0.4);

\draw[black] (10,3-2-0.7) circle (0.4); \fill[gray3] (10,3-2-0.7) circle (0.4);
\draw[black] (11-0.4,3-2-0.8) circle (0.4); \fill[gray1] (11-0.4,3-2-0.8) circle (0.4);
\draw[black] (12+0.5,3-2-0.5) circle (0.4); \fill[gray4] (12+0.5,3-2-0.5) circle (0.4);
\draw[black] (13,3-2-0.5-0.2) circle (0.4); \fill[gray2] (13,3-2-0.5-0.2) circle (0.4);

\draw[black] (10,3-3-0.5+0.3) circle (0.4); \fill[gray5] (10,3-3-0.5+0.3) circle (0.4);
\draw[black] (11-0.3,3-3-0.5+0.3) circle (0.4); \fill[gray2] (11-0.3,3-3-0.5+0.3) circle (0.4);
\draw[black] (12+0.4,3-3-0.5+0.3) circle (0.4); \fill[gray3] (12+0.4,3-3-0.5+0.3) circle (0.4);
\draw[black] (13-0.2,3-3-0.5+0.3) circle (0.4); \fill[gray1] (13-0.2,3-3-0.5+0.3) circle (0.4);

\draw[black] (14.5, -1) rectangle (18.5, 3);

\draw[black] (16-0.2,3-1-0.5+0.8) circle (0.4); \fill[gray4] (16-0.2,3-1-0.5+0.8) circle (0.4);


\draw[black] (16-0.5,3-3-0.5+0.5) circle (0.4); \fill[gray2] (16-0.5,3-3-0.5+0.5) circle (0.4);
\draw[black] (18-0.2,3-3-0.5+0.5) circle (0.4); \fill[gray1] (18-0.2,3-3-0.5+0.5) circle (0.4);

\node at (1.5, -1.4) {\Large$t_1$};
\node at (6.6, -1.4) {\Large$t_2$};
\node at (11.5, -1.4) {\Large$t_3$};
\node at (16.5, -1.4) {\Large$t_4$};

\draw[red,->, line width=0.6mm] (3.6,1) -- (4.4,1);
\draw[red,->, line width=0.6mm] (8.6,1) -- (9.4,1);
\draw[red,->, line width=0.6mm] (13.6,1) -- (14.4,1);

\end{tikzpicture}}
\caption{ Representation of the evolution of pixels as they tend to aggregate in different clusters. 
}
\label{fig:aggregation}
\end{figure}
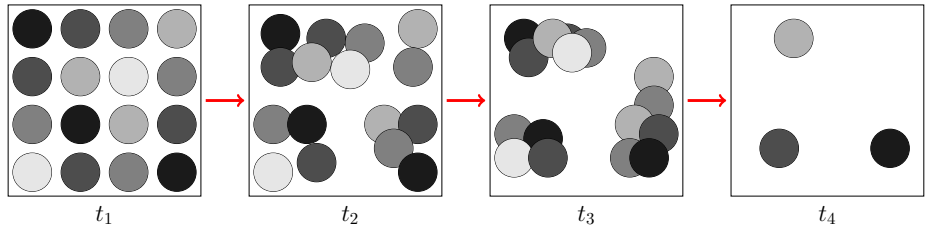

An application of the Hegselman-Krause model for data-clustering problems has been proposed in \cite{HPV}. The idea is to extend the 2D model by characterising each particle with an internal feature $c_{i} \in [0,1]$ that represents the gray color of the $i$th pixel. Therefore, we interpret each pixel in the image as a particle characterized by a position vector and the static feature $c$ as shown in Figure \ref{ISHK}. 
To address the segmentation task, we can define a dynamic feature for the system of pixels through an interaction function that accounts for alignment processes among pixels with sufficiently similar features. In particular, let us consider the following 
\begin{equation} \label{int.func}
    P_{\Delta_{1},\Delta_{2}}(\mathbf{x}_{i},\mathbf{x}_{j},c_{i},c_{j}) =  \chi(|\mathbf{x}_{i} - \mathbf{x}_{j}| \leq \Delta_{1})\chi(|c_{i} - c_{j}| \leq \Delta_{2}). 
\end{equation}
Therefore, the time-continuous evolution for the system of pixels is given by
\begin{equation} \label{tcdynamics}
\begin{split}
        \frac{d}{dt}\mathbf{x}_{i} &= \frac{1}{N}\sum_{j=1}^{N} P_{\Delta_{1},\Delta_{2}}(\mathbf{x}_{i},\mathbf{x}_{j},c_{i},c_{j})(\mathbf{x}_{j} - \mathbf{x}_{i}) \\
        \frac{d}{dt}c_{i} &= 0
\end{split}
\end{equation}
 In this case we introduced two confidence bounds $\Delta_{1} \geq 0$, $\Delta_{2} \geq 0$ taking into account the position and the gray level of the pixels, respectively. In this way, the interactions between the pixels will generate a large time distribution which is characterized by several clusters, depending on the value of $\Delta_{1}$ and $\Delta_{2}$. Hence, coherently with k-means methods, see e.g. \cite{MPSKPM}, a pixel belongs to a cluster $\mathcal C_{\mu} = \{\mathbf x_i : \| \mathbf x_i-\mu\|\le \alpha\}$, being $\alpha>0$ the pixel size, if it is sufficiently close to the local quantity $\mu \in \mathbb R^2$. We highlight how we are only interested in clustering with respect to the space variable. This dynamics is represented in Figure \ref{fig:aggregation}. 

Biomedical images are often subject to ambiguities arising from various sources of uncertainty related to clinical factors and potential bottlenecks in data acquisition processes \cite{BAJT,KWKP}. These uncertainties can be broadly categorized into aleatoric uncertainty, stemming from inherent stochastic variations in the data collection process, and epistemic uncertainty, relates to uncertainties in model parameters and can lead to deviations in the results. Aleatoric uncertainties poses significant challenges in  image segmentation, as image processing models must contend with limitations in the raw acquisition data. Addressing these uncertainties is critical, and the study of uncertainty quantification in image segmentation is an expanding field aimed at developing robust segmentation algorithms capable of mitigating erroneous outcomes. To this end, in~\cite{CPLFZ}, it has been proposed an extension of \eqref{tcdynamics} to consider segmentation of biomedical images. In particular, the particle model \eqref{tcdynamics_new} has been integrated a nonconstant stochastic part to take into account aleatoric uncertainties arising from the data acquisition process. These uncertainties may include factors such as motion artifacts or field inhomogeneities in magnetic resonance imaging (MRI). They modified equation~\ref{tcdynamics} as follows:
\begin{equation} \label{tcdynamics_new}
\begin{split}
        d\mathbf{x}_{i} &= \frac{1}{N}\sum_{j=1}^{N} P_{\Delta_{1},\Delta_{2}}(\mathbf{x}_{i},\mathbf{x}_{j},c_{i},c_{j})(\mathbf{x}_{j} - \mathbf{x}_{i})dt  + \sqrt{2 \sigma^2 D(c)} d\mathbf{W}_{i} \\
        \frac{d}{dt}c_{i} &= 0
\end{split}
\end{equation}
where $\{\mathbf{W}_i\}_{i=1}^N$ is set of independent Wiener processes, $P_{\Delta_1,\Delta_2}(\cdot,\cdot,\cdot,\cdot)\in[0,1]$ is the interaction function defined in \eqref{int.func}, and $D(c)\ge0$ quantifies the impact of diffusion related to the value of the feature $c \in[0,1]$.   Since the aleatoric uncertainties are expected to appear far away from the static feature's boundaries, only diffusion functions that are maximal at the center and satisfy $D(0) = D(1) = 0$ are considered. Similarly to \eqref{bi.scheme}, we may introduce the following binary interaction scheme by writing $(\mathbf{x},\mathbf{x}_*):= (\mathbf{x}_i(t),\mathbf{x}_j(t))$ a random couple of pixels having features $(c,c_*):= (c_i(t),c_j(t))$. We get
\begin{equation} \label{bi.scheme_c}     
\begin{split} 
        \mathbf{x'} &= \mathbf{x} + \epsilon P_{\Delta_{1},\Delta_{2}}(\mathbf{x},\mathbf{x_{*}},c,c_{*})(\mathbf{x_{*}} - \mathbf{x}) + \sqrt{2\sigma^2  D(c)} \eta \\
        \mathbf{x_{*}'} &= \mathbf{x_{*}} + \epsilon P_{\Delta_{1},\Delta_{2}}(\mathbf{x_{*}},\mathbf{x},c_{*},c)(\mathbf{x} - \mathbf{x_{*}}) + \sqrt{2\sigma^2 D(c_{*})} \eta \\
        c_{*}' &= c_{*}\\
        c' &= c, 
\end{split} 
\end{equation}
where $(\mathbf{x}^\prime,\mathbf{x}_*^\prime):= (\mathbf{x}_i(t+\Delta t),\mathbf{x}_j(t+\Delta t))$ and $(c^\prime,c_*^\prime):= (c_i(t+\Delta t),c_j(t+\Delta t))$. At the statistical level, as in \cite{CPLFZ}, we may follows the approach described in Section \ref{sect:consensus_kin}. Hence, we introduce the distribution function $f = f(\mathbf{x},c,t) : R^{2} \times [0,1] \times R_{+} \rightarrow R_{+}$, such that $f (\mathbf{x}, t)d\mathbf{x}$ represents the fraction of agents/particles in $[x_1, x_1 + dx_1) \times [x_2, x_2 + dx_2]$ characterized by a feature $c \in [0,1]$ at time $t\ge0$. The evolution of $f$ whose interaction follow the binary scheme \eqref{bi.scheme_c} is given by the following Boltzmann-type equation
\begin{equation}\label{eq:boltzmannc}
\begin{split}
    &\dfrac{d}{dt} \int_0^1\int_{\mathbb R^2}\varphi(\mathbf{x},c)f(\mathbf{x},c,t)d\mathbf{x}dc  = \\
    &\quad  \left \langle \int_{[0,1]^2}\int_{\mathbb R^{4}} (\varphi(\mathbf{x}^\prime,c) - \varphi(\mathbf{x},c)) f(\mathbf{x},c,t)f(\mathbf{x}_*,c_*,t)d\mathbf{x} d\mathbf{x}_{*}\,dc\,dc_* \right \rangle,
\end{split}
\end{equation}  
Hence, since the feature is not evolving in time, we can proceed as in Section \ref{sect:consensus_kin} to derive in the quasi-invariant limit for $\epsilon\to 0^+$ the corresponding Fokker-Planck-type PDE

\begin{equation} \label{FP2}
\begin{split}
    \partial_{t} f(\mathbf{x},c,t) = \nabla_{x} \cdot \Big [\Xi[g]_{\Delta_{1},\Delta_{2}}(\mathbf{x},c,t)f(\mathbf{x},c,t) + \sigma^2 D(c)\nabla_{x}f(\mathbf{x},c,t)\Big ]    
\end{split}
\end{equation}
where 
\[
\Xi[g]_{\Delta_{1},\Delta_{2}}(\mathbf{x},c,t) = \int_0^1\int_{\mathbb R^2}P_{\Delta _1,\Delta_2}(\mathbf{x},\mathbf{x}_*,c,c_*)(\mathbf{x}-\mathbf{x}_*)f(\mathbf{x}_*,c_*,t)d\mathbf{x}_*\,dc_*. 
\]

\section{Evaluation metrics and parameters estimation}\label{sect:metrics}

In this section we present classical Direct Simulation Monte Carlo (DSMC) methods to numerically approximate the evolution of \eqref{eq:boltzmannc} as quasi-invariant approximation of the Fokker-Planck equation \eqref{FP2}. The resulting numerical algorithm is fundamental to estimate consistent parameters from MRI images. To this end, we present several loss metrics with the aim to compare the result of our model-based approach with existing methods for biomedical image segmentation. In this work we focus exclusively on binary metrics. For evaluation of segmentation with multiple labels we point the reader to \cite{TH} for a detailed presentation of various metrics.

\subsection{DSMC algorithm for image segmentation}\label{sect:DSMC}

The numerical approximation of Boltzmann-type equations has been deeply investigated in the recent decades, see e.g. \cite{DP2014,PR}. The approximation of this class of equations is particularly challenging due to  the curse of dimensionality brought up by the multidimensional integral of the collision operator, and the presence of multiple scales. Furthermore, the preservation of relevant physical quantities are essential for a correct description of the underlying physical problem \cite{PZ}. 

In view of its computational efficiency, in the following we will adopt a DSMC approach. Indeed the computational cost of this method is $O(N)$ where $N$ is the number of particles. Next, we describe the DSMC method based on a Nanbu-Bavosky scheme \cite{PR}. We begin by randomly selecting $N/2$ pairs of particles and making them evolve following the binary scheme presented in \eqref{bi.scheme}. We consider a time interval $[0,T]$ which we divide in $N_{t}$ intervals of size $\Delta t>0$. The DSMC approach for the introduced kinetic equation is based on a first order forward time discretization. In the following we will always consider the case $\Delta t= \epsilon>0$ such that all the particles are going to interact, see \cite{PR} for more details. We introduce the stochastic rounding of a positive real number x as:

\begin{equation}
    Sround(x) = 
    \begin{cases}
      \lfloor x \rfloor + 1 \hspace{0.5cm}\text{with probability} \hspace{0.5cm}x - \lfloor x \rfloor\\
      \lfloor x \rfloor \hspace{1.13cm}\text{with probability} \hspace{0.5cm}1 - x + \lfloor x \rfloor 
    \end{cases} 
\end{equation}

where $\lfloor x \rfloor$ is the integer part of $x$. The random variable $\eta$ is sampled from a 2D Gaussian Distribution centered at zero and a diagonal covariance matrix. 

\begin{algorithm}[ht] 
    \caption{DSMC algorithm for Boltzmann equation}\label{alg:Boltz}
    \begin{algorithmic}[1]
        \State Given $N$ particles $(\textbf{x}_n^0 , c_n^0)$, with $n = 1, \dots, N$ computed from the initial distribution $f_0(\textbf{x}, c)$;
        \For{$t=1$ \textbf{to}  $N_t$}
        \State set $n_p=\textrm{Sround}(N/2)$;
        \State sample $n_p$ pairs $(i,j)$ uniformly without repetition among all possible pairs of particles at time step $t$;
        \State for each pair $(i,j)$, sample $\eta$,$\eta_*$
        \State for each pair $(i,j)$, compute the data change:
        \begin{equation}
            \begin{aligned}
               \Delta\mathbf{x}_i^t &= \epsilon P_{\Delta_1, \Delta_2}(\mathbf{x}_i^t, \mathbf{x}_j^t, c_i^0, c_j^0) (\mathbf{x}_j^t-\mathbf{x}_i^t) + \sqrt{2\sigma^2 D(c_i^0)} \boldsymbol{\eta} \\
               \Delta\mathbf{x}_j^t &= \epsilon P_{\Delta_1, \Delta_2}(\mathbf{x}_j^t, \mathbf{x}_i^t, c_j^0, c_i^0) (\mathbf{x}_i^t-\mathbf{x}_j^t) + \sqrt{2\sigma^2 D(c_j^0)} \boldsymbol{\eta}_*
            \end{aligned}
        \end{equation}
        compute
            \begin{equation}
               \mathbf{x}_{i,j}^{t+1} = \mathbf{x}_{i,j}^t + \Delta \mathbf{x}_{i,j}^t 
            \end{equation}
        \EndFor
    \end{algorithmic}
\end{algorithm}

\subsection{Generation of a model-oriented segmentation masks}
In this section we present the procedure to estimate the Segmentation Mask of Brain Tumor Images. The procedure described in this section closely follows the methodology presented in ~\cite{CPLFZ}. For a given image, we define the feature's values in relation to the gray level of each pixel. In more detail, for a given pixel $i  \in \{1,\dots,N\}$ we define
\[
c_i = \dfrac{C_i - \min_{i = 1,\dots,N}{C_i}}{\max_{i = 1,\dots,N}{C_i} - \min_{i = 1,\dots,N}{C_i}} \in [0,1],
\]
being $C_i$, $i = 1,\dots,N$, the gray value of the original image. Therefore, the value $c_i = 1$ represents a white pixel and $c_i = 0$ represents black pixel. 
 In particular, for this work we used the \textit{brain tumor dataset} that consists of 3D in multi-parametric MRI of patients affected by glioblastoma or lower-grade glioma, publicly available in the context of the Brain Tumor Image Segmentation Challenge \url{http://medicaldecathlon.com/}. The acquisition sequences include $T_1$-weighted, post-Gadolinium contrast $T_1$-weighted, $T_2$-weighted and $T_2$ Fluid-Attenuated Inversion Recovery volumes. Each MRI scan is accompanied by corresponding ground truth segmentation mask, which is a binary image where anatomical regions of interest are highlighted as white pixels while all other areas are represented as black pixels. These ground truth segmentation masks were manually delineated by experienced radiologists and specifically identify three structures: "tumor core", "enhancing tumor" and "whole tumor". We evaluate the performances of the DSMC algorithm for two different segmentation tasks: the "tumor core" and the "whole tumor" annotations. For the first task we use a single slice in the axial plane of the post-Gadolinium contrast $T_1$-weighted scans while for the second task we use a single slice in the axial plane of the $T_2$-weighted scans. The procedure to generate the segmentation masks is as follows:

\begin{enumerate}
    \item We begin by associating each pixel with a position vector $(x_{i},y_{i})$ and with static feature $c_{i}$. We scale the vector position to a domain $[-1,1]\times[-1,1]$ and the static feature to $[0,1]$.
    \item We apply a DSMC approach as described in  Algorithm~\ref{alg:Boltz} to numerically approximate the large-time solution of the Boltzmann-type model defined  in~\eqref{eq:boltzmann}. This approach enables pixels to aggregate into clusters based on their Euclidean distance and gray color level. 
    \item The segmentation masks are generated by assigning to the original position of each pixel the mean value of the clusters they belong to. In this way we generate a multi-level mask composed of a number of homogenous regions. 
    \item Finally we obtain the binary mask by defining a threshold $\Tilde{c}$ such that:
    \begin{equation}
        c_{i} = 
        \begin{cases}
            1 \ if \  c \geq \Tilde{c} \\
            0 \ if \  c < \Tilde{c}
        \end{cases}
    \end{equation}
    For all the following experiments, $\Tilde{c}$ is defined as the 10th percentile of pixels in the image that belong to the region of interest. This percentile was chosen as an optimal value for brain tumor images; however, it could also be considered as a parameter to be optimized within the process outlined in Section~\ref{Par.Opt}.
\end{enumerate}

\begin{figure}[H]
    \centering
    \includegraphics[width=1.\linewidth]{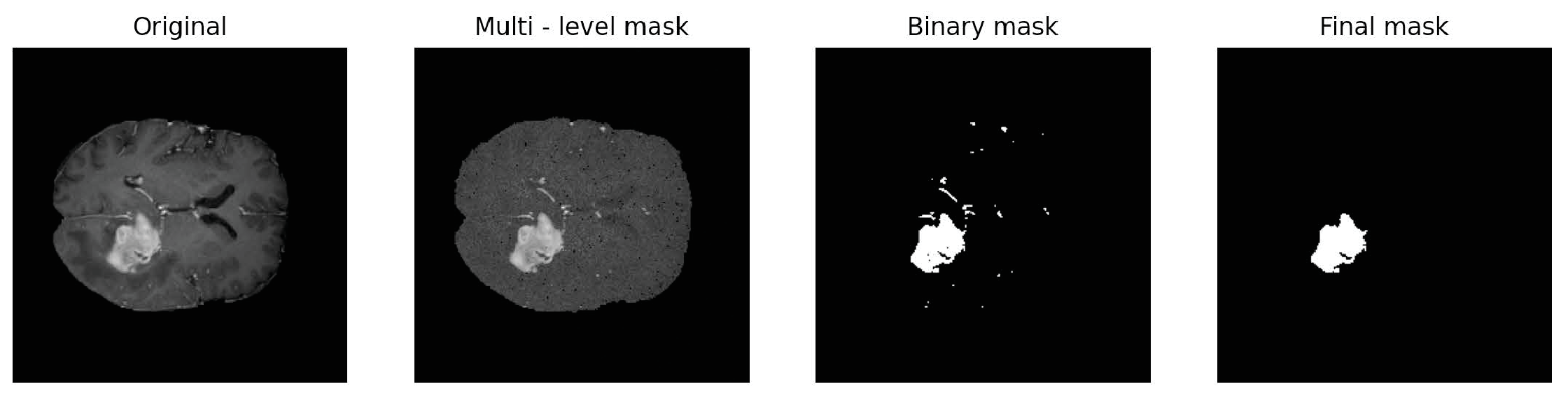}
    \caption{ Summary of the segmentation process. The first image shows the input image. By means of the Algorithm \ref{alg:Boltz} we generate the Multi-level mask where we reassign each picture's gray level to the mean value of the cluster they are assigned to. The binary mask is produced as result of the binarization process. And the Final mask is the result after the two morphological refinements steps have been applied. }
    \label{fig:procedure}
\end{figure}

Following this procedure, we apply two morphological refinement steps to remove small regions that have been misclassified as foreground parts and to fill small regions that have been incorrectly categorized as background pixels. We begin by labeling all the connected pixels in the foreground and reassigning them to the background those whose number of pixels is less than a certain threshold. Then we repeat the same procedure but for the pixels in the background. To this end, we use the scikit-image python library that detects distinct objects of a binary image \cite{VDWSNIBWYGY}. This allows us to obtain more precise segmentation masks by reducing small imperfections. This entire process is illustrated in Figure \ref{fig:procedure}. 

\subsubsection{Parameters optimization} \label{Par.Opt}
In this section, we outline the procedure for optimizing the parameters $\Delta_{1} > 0$, $\Delta_{2} > 0$ and $\sigma^{2} > 0$ that best approximate the ground truth segmentation masks. The goal is to identify the parameter configuration that minimizes the discrepancy between the computed and ground truth masks, measured through a predefined loss metric.
To achieve this, we solve the following minimization problem:
    \begin{equation}
        \displaystyle \min_{\Delta_{1},\Delta_{2},\sigma^{2} > 0} Loss(S_{g}, S_{t}) = \min_{\Delta_{1},\Delta_{2},\sigma^{2} > 0} {1-Metric(S_{g}, S_{t})}
        \label{optimization}
    \end{equation}

where $S_{g}$ is the Ground Truth segmentation mask and $S_{t}$ is the segmentation mask computed by the model. The different $Loss$ metrics quantify the discrepancy between the masks, with lower values indicating greater similarity. Accordingly, the $Metric$ function, detailed in Section~\ref{sec:metrics}, measures the similarity between the two masks, with higher values indicating better agreement. The relationship $Loss = 1- Metric$ is satisfied when the $Metric$ is defined to take a value of 1 for perfect agreement and 0 for complete mismatch.

To solve the optimization problem~\eqref{optimization}, we used the \textit{Hyperopt} package~\cite{BKEYC}. This optimization method randomly samples the parameter configurations from predefined distributions and selects the configuration that minimizes the $Loss$ metric. This sampling process is repeated for a predefined number of iterations. In this work, we sample the values of our parameters from the following distributions:
\begin{equation}
\begin{gathered}
    \Delta_{1} \sim U(\Delta x,0.7) \\
    \Delta_{2} \sim U(0.05 ,0.3) \\
    \sigma^{2} \sim \text{log-uniform}(e^{-5},1)
\end{gathered}
\end{equation}
where $\Delta x$ represents the distance between the initial positions of the pixels at $t=0$. We perform 300 iterations of the optimization process. To ensure reproducibility and correctly compare the different results obtained, the random seed for parameter sampling is fixed. 

\subsection{Segmentation Metrics}\label{sec:metrics}
Next, we introduce the principal optimization metrics used for evaluating a binary segmentation mask. We define $\{S_{g}^{0},S_{g}^{1},S_{t}^{0},S_{t}^{1}\}$ where $S_{g}^{0}$ and $S_{g}^{1}$ represent the set of pixels that belong to the background and foreground of the ground truth segmentation mask respectively. Same applies for ${S_{t}^{0},S_{t}^{1}}$ but for the binary mask we want to evaluate. One could also wish to asses the validity of a segmentation mask with multiple labels, we refer to \cite{TH} for an introduction to the subject. Figure~\ref{fig:metrics} presents a summary of the key terms used in the definitions of metrics.

\colorlet{shape edge}{blue!50}
\colorlet{shape area}{blue!20}

\tikzset{filled/.style={fill=shape area, draw=shape edge, thick}, outline/.style={draw=shape edge, thick}}

\setlength{\parskip}{5mm}

\begin{figure}[!ht]
    \centering
    \begin{subfigure}[c]{0.3\textwidth}
        \begin{tikzpicture}
            \begin{scope}
                \clip plot [smooth cycle, tension=0.7] coordinates {(0,0) (1,1.5) (2,1) (3,0) (2,-1) (1,-1.5) (0,-1)};
                \fill[filled] plot [smooth cycle, tension=0.7] coordinates {(0.5,0) (1.2,1.2) (2,1) (3,0) (2,-1) (1.5,-1.5) (1,-1)}; 
            \end{scope}
            \draw[outline] plot [smooth cycle, tension=0.7] coordinates {(0,0) (1,1.5) (2,1) (3,0) (2,-1) (1,-1.5) (0,-1)};
            \node at (0.3, -0.7) {$S_t^1$}; 
            \draw[outline] plot [smooth cycle, tension=0.7] coordinates {(0.5,0) (1.2,1.2) (2,1) (3,0) (2,-1) (1.5,-1.5) (1,-1)};
            \node at (1.7, -0.7) {$S_g^1$};
            \node[anchor=south] at (current bounding box.north) {$S_t^1 \cap S_g^1$ (TP)};
        \end{tikzpicture}
        \caption{Intersection area or true positive (TP).}
    \end{subfigure}
    \hspace{1.5cm}
    \begin{subfigure}[c]{0.3\textwidth}
        \begin{tikzpicture}
            \draw[filled] plot [smooth cycle, tension=0.7] coordinates {(0,0) (1,1.5) (2,1) (3,0) (2,-1) (1,-1.5) (0,-1)} 
                         plot [smooth cycle, tension=0.7] coordinates {(0.5,0) (1.2,1.2) (2,1) (3,0) (2,-1) (1.5,-1.5) (1,-1)};  
            \node at (0.3, -0.7) {$S_t^1$};
            \node at (1.7, -0.7) {$S_g^1$};
            \node[anchor=south] at (current bounding box.north) {$S_t^1 \cup S_g^1$};
        \end{tikzpicture}
        \caption{Union area.}
    \end{subfigure}
    \vskip 1cm
    \begin{subfigure}[c]{0.3\textwidth}
        \begin{tikzpicture}
            \begin{scope}
                \clip plot [smooth cycle, tension=0.7] coordinates {(0,0) (1,1.5) (2,1) (3,0) (2,-1) (1,-1.5) (0,-1)};
                \draw[filled, even odd rule] plot [smooth cycle, tension=0.7] coordinates {(0,0) (1,1.5) (2,1) (3,0) (2,-1) (1,-1.5) (0,-1)}
                                             plot [smooth cycle, tension=0.7] coordinates {(0.5,0) (1.2,1.2) (2,1) (3,0) (2,-1) (1.5 ,-1.5) (1,-1)};
            \end{scope}
            \draw [outline] plot [smooth cycle, tension=0.7] coordinates {(0,0) (1,1.5) (2,1) (3,0) (2,-1) (1,-1.5) (0,-1)};
            \node at (0.3, -0.7) {$S_t^1$};
            \draw[outline] plot [smooth cycle, tension=0.7] coordinates {(0.5,0) (1.2,1.2) (2,1) (3,0) (2,-1) (1.5,-1.5) (1,-1)};
            \node at (1.7, -0.7) {$S_g^1$};
            \node[anchor=south] at (current bounding box.north) {$S_t^1 - S_g^1$ (FP)};
        \end{tikzpicture}
        \caption{False positive (FP).}
    \end{subfigure}
    \hspace{2.3cm}
    \begin{subfigure}[c]{0.3\textwidth}
        \begin{tikzpicture}
            \begin{scope}
                \clip plot [smooth cycle, tension=0.7] coordinates {(0.5,0) (1.2,1.2) (2,1) (3,0) (2,-1) (1.5,-1.5) (1,-1)};
                \draw[filled, even odd rule] plot [smooth cycle, tension=0.7] coordinates {(0,0) (1,1.5) (2,1) (3,0) (2,-1) (1,-1.5) (0,-1)}
                                             plot [smooth cycle, tension=0.7] coordinates {(0.5,0) (1.2,1.2) (2,1) (3,0) (2,-1) (1.5,-1.5) (1,-1)};
            \end{scope}
            \draw[outline] plot [smooth cycle, tension=0.7] coordinates {(0,0) (1,1.5) (2,1) (3,0) (2,-1) (1,-1.5) (0,-1)};
            \node at (0.3, -0.7) {$S_t^1$};
            \draw[outline] plot [smooth cycle, tension=0.7] coordinates {(0.5,0) (1.2,1.2) (2,1) (3,0) (2,-1) (1.5,-1.5) (1,-1)};
            \node at (1.7, -0.7) {$S_g^1$};
            \node[anchor=south] at (current bounding box.north) {$S_g^1 - S_t^1$ (FN)};
        \end{tikzpicture}
        \caption{False negative (FN).}
    \end{subfigure}
    \vskip 1cm
    \begin{subfigure}[c]{0.3\textwidth}
        \begin{tikzpicture}
            \draw[outline] plot [smooth cycle, tension=0.7] coordinates {(0,0) (1,1.5) (2,1) (3,0) (2,-1) (1,-1.5) (0,-1)};
            \node at (0.3, -0.7) {$S_t^1$};
            \draw[outline] plot [smooth cycle, tension=0.7] coordinates {(0.5,0) (1.2,1.2) (2,1) (3,0) (2,-1) (1.5,-1.5) (1,-1)};
            \node at (1.7, -0.7) {$S_g^1$};
            \draw[blue, line width=0.6mm] plot [smooth, tension=0.7] coordinates {(2,1) (3,0) (2,-1)};
        
            \node[anchor=south] at (current bounding box.north) {$B_t^{\tau=0} \cap B_g^{\tau=0}$};
        \end{tikzpicture}
        \caption{Intersection of boundaries at $\tau = 0$.}
    \end{subfigure}
    \hspace{1.5cm}
    \begin{subfigure}[c]{0.3\textwidth}
        \begin{tikzpicture}
 \draw[outline] plot [smooth cycle , tension=0.7] coordinates {(0,0) (1,1.5) (2,1) (3,0) (2,-1) (1,-1.5) (0,-1)};
            \node at (0.3, -0.7) {$S_t^1$};
            \draw[outline] plot [smooth cycle, tension=0.7] coordinates {(0.5,0) (1.2,1.2) (2,1) (3,0) (2,-1) (1.5,-1.5) (1,-1)};
            \node at (1.7, -0.7) {$S_g^1$};
            \draw[blue, line width=0.9mm] plot [smooth, tension=0.7] coordinates {(1.5,1.3) (2,1) (3,0) (2,-1) (1.58,-1.5)}; 
            \draw[blue, line width=0.9mm] plot [smooth, tension=0.7] coordinates {(1.4,1.52) (2,1) (3,0) (2,-1) (1.52,-1.35)}; 

            \node[anchor=south] at (current bounding box.north) {$B_t^{\tau>0} \cap B_g^{\tau>0}$};
        \end{tikzpicture}
        \caption{Intersection of boundaries at $\tau>0$.}
    \end{subfigure}

\caption{Representation of the relevant areas between the predicted $S_t^1$ and ground truth $S_g^1$ segmentation masks. $B_t^{\tau}$ and $B_g^{\tau}$ represent the corresponding boundaries with a $\tau$ threshold.}
\label{fig:metrics}
\end{figure}
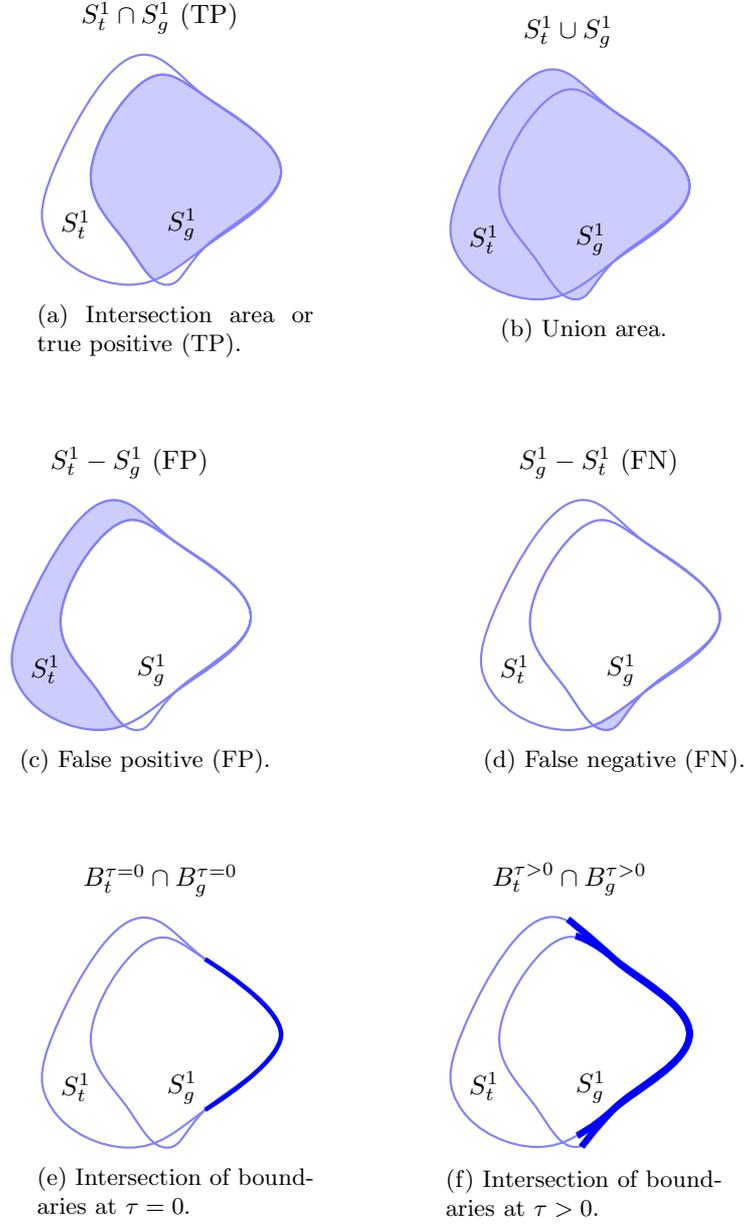

\subsubsection{Volumetric  and Surface Dice Indexes}

The \textit{Volumetric Dice} index, also known as the Standard Volumetric Dice Similarity Coefficient, first introduced in \cite{DI}, is the most used metric when evaluating volumetric segmentations.  It is defined as follows:

\begin{equation} \label{DICE}
    \textrm{DICE} = \frac{2 |S_{g}^{1} \cap S_{t}^{1}|}{|S_{g}^{1}| + |S_{t}^{1}|}
\end{equation}

where, $|\cdot|$ indicates the total number of pixels of the considered region.
This metric is equal to one if there is a perfect overlap between the two segmentation masks and null if both segmentations are completely disjoint. Since the Volumetric Dice coefficient is the most commonly used metric for segmentations, especially in the biomedical field, the results are highly interpretable and can be compared with those obtained in other studies. However, when assessing surface segmentation masks, the Volumetric Dice coefficient can yield suboptimal results. This limitation arises because the Volumetric Dice coefficient evaluates the similarity between segmentation masks based on pixel overlap without considering the spatial accuracy of the boundaries. Specifically, it treats all pixel displacements equally, without considering how far a segmentation error might be from the true boundary of the object. This means that segmentations with minor errors spread across multiple areas and those with a major error in a single area might receive similar scores. To address this limitation, the Surface Dice Similarity Coefficient was presented in \cite{N_etal} as a metric that can assess the accuracy of segmentation masks by considering the similarity of their boundaries. We define $\zeta : I \rightarrow R^{2}$ as a parameterization of $\partial S_{i}$, the boundary of the segmentation mask $S_{i}$. The border region $B^{(\tau)}_{i}$, which is a region around the boundary $\partial S^{i}$ with tolerance $\tau$, is defined as:
\begin{equation}
    B^{(\tau)}_{i} = \Big \{ x \in R^{2} / \ \exists \  y \in I \ s.t. || x  - \zeta(y) || \leq \tau \Big \}
\end{equation}
where, $\tau$ is a positive real number that defines the maximum allowable distance from the boundary $\partial S^{i}$ for a point $x$ to be considered part of the border region $B^{(\tau)}_{i}$. 
The Surface Dice Similarity Coefficient between $S_{t}$ and $S_{g}$ with tolerance $\tau$ is defined as:

\begin{equation}
    R^{(\tau)}_{g,t} = \frac{2 \left| B_{g}^{(\tau)} \cap B_{t}^{(\tau)}  \right|}{\left| B_{g}^{(\tau)}  \right| + \left| B_{t}^{(\tau)}  \right|}\
\end{equation}

$R^{(\tau)}_{g,t}$, ranges from 0 to 1. A score of 1 indicates a perfect overlap between the two surfaces, while a score of 0 indicates no overlap. A larger value of $\tau$ results in a wider border region, making the metric more tolerant to small deviations in the boundary.

\subsubsection{Jaccard Index}
The Jaccard Index (JAC) \cite{J}, similar to the Volumetric Dice coefficient, measures the similarity between two segmentations by quantifying the overlap between the computed mask and the ground truth. It is defined as the ratio between the intersection and the union of the foreground's segmentation masks

\begin{equation} \label{JAC}
    \textrm{JAC} = \frac{|S_{g}^{1} \cap S_{t}^{1}|}{|S_{g}^{1} \cup S_{t}^{1}|}. 
\end{equation}
The JAC Index and the Volumetric Dice coefficient are closely related since we have
\begin{equation} \label{dice jac rel}
\begin{gathered}
     \textrm{JAC} = \frac{\textrm{DICE}}{2 - \textrm{DICE}} \qquad 
     \textrm{DICE}  = \frac{2 \textrm{JAC}}{1 + \textrm{JAC}}. 
\end{gathered}
\end{equation}
From \eqref{dice jac rel} we get the relationship between the JAC index and the Volumetric Dice coefficient. While both are widely used for measuring segmentation similarity, they can produce slightly different results. To understand the implications of these differences, we can analyze how their absolute and relative errors are related. 

\begin{Definition}[Absolute Approximation]
    A similarity S is absolutely approximated by $\Tilde{S}$ with error $\epsilon \geq 0$ if the following holds for all y and $\Tilde{y}$: \vspace{0.5cm} \\
    \centering
    $|S(y,\Tilde{y}) - \Tilde{S}(y,\Tilde{y})| \leq \epsilon$
\end{Definition}

\begin{Definition}[Relative Approximation]
    A similarity S is relatively approximated by $\Tilde{S}$ with error $\epsilon \geq 0$ if the following holds for all y and $\Tilde{y}$:\vspace{0.5cm} \\
    \centering
    $\dfrac{\Tilde{S}(y,\Tilde{y})}{1 + \epsilon} \leq S(y,\Tilde{y}) \leq \Tilde{S}(y,\Tilde{y})\cdot(1 + \epsilon).$
\end{Definition}
The following  result holds
\begin{Proposition}
    JAC and Volumetric Dice approximate each other with a relative error of 1 and an absolute error of $3 - 2\sqrt{2}$.
\end{Proposition}

We point the reader to \cite{BEBVMBB} for a deeper comparison between the Jaccard and Volumetric Dice Index.

\subsubsection{F-measure}

The $F_{\beta}-$measure is commonly used as an information retrieval metric~\cite{Sasaki,C}. To define this metric, we first introduce two terms: Positive Predicted Value (PPV) and True Positive Rate (TPR), which are also known as Precision and Sensitivity, respectively. The \textit{Precision} metric quantifies the proportion of correctly predicted foreground pixels (true positives, TP) out of all pixels predicted as foreground (TP + false positives, FP). The \textit{Sensitivity} measures the proportion of actual foreground pixels (TP) correctly identified by the model out of all actual foreground pixels (TP + false negatives, FN). These two metrics can be expressed as follows

\begin{equation}
\begin{split}\label{eq:precision_sensitivity}
   \textrm{ Precision} = \textrm{PPV} = \frac{\textrm{TP}}{\textrm{TP + FP}} \\
    \textrm{Sensitivity = TPR } = \frac{\textrm{TP}}{\textrm{TP + FN}}
\end{split}
\end{equation}

The \textit{Precision} metric indicates how many of the predicted foreground pixels are actually correct. 
The \textit{Sensitivity}, on the other hand, measures how many of the actual foreground pixels were correctly predicted by the model.

We can define the $F_{\beta}-$measure as a combination of Precision and Sensitivity, with a parameter $\beta$ that controls the trade-off between these two metrics. Specifically, the $F_{\beta}-$measure is given by

\begin{equation} \label{F_measure_rel}
\textrm{FMS}_{\beta} = \frac{(\beta^{2}+1) \cdot \text{PPV} \cdot \text{TPR}}{\beta^{2} \cdot \text{PPV} + \text{TPR}}
\end{equation}
We may observe that if $\beta = 1$ we obtain the Volumetric Dice metric.

To understand the impact of $\beta$ in the $F_{\beta}-$measure, we can substitute the definitions of PPV and TPR into \eqref{F_measure_rel}, which results in the following
\begin{equation}
   \textrm{FMS}_{\beta} = \frac{(\beta^{2} + 1) \textrm{TP}^{2}}{(\beta^{2} + 1) \textrm{TP}^{2} + \textrm{TP} (\beta^{2}\textrm{FN + FP})}
\end{equation}

If $\beta > 1$ the $F_{\beta}-$measure emphasizes minimizing False Negatives (maximizing Sensitivity), which can lead to more False Positives (lower Precision).
If $\beta < 1$ the $F_{\beta}-$measure focuses on minimizing False Positives (maximizing Precision), potentially increasing the number of False Negatives (lower Sensitivity).

Furthermore, it can be noticed that 

\begin{equation}
\lim_{\beta\to\infty} \frac{(\beta^{2} + 1) \textrm{TP}^{2}}{(\beta^{2} + 1)\textrm{TP}^{2} + \textrm{TP} (\beta^{2}\textrm{FN + FP})} = \textrm{Sensitivity} = \frac{\textrm{TP}}{\textrm{TP + FN}}
\end{equation}

since for $\beta\gg0$ we neglect the contribution of the False Positives by considering only the contribution of the False Negatives where we re-obtain the TPR metrics defined in \eqref{eq:precision_sensitivity}. 

In summary, thanks to the ${\beta}$ parameter, the $F_{\beta}$-measure offers a flexible way to evaluate segmentation models by allowing for a tunable balance between Precision and Sensitivity. It provides a useful metric when dealing with class imbalances, especially in the field of medical imaging, where the relative importance of false positives and false negatives can vary according to each segmentation task.

\section{Numerical Results}\label{sect:results}

\subsection{Impact of different diffusion functions}

\begin{figure}
    \centering
    \includegraphics[width=1.05\linewidth]{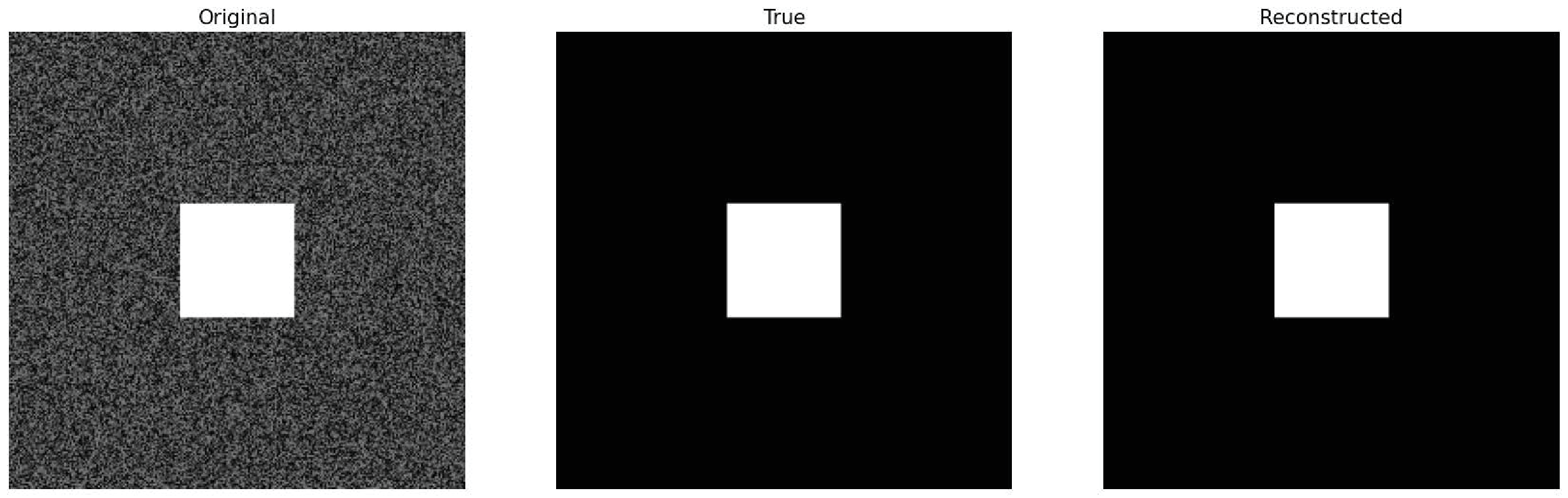} 
    \caption*{\centering (a)}
    \vspace{0.3cm}
    \centering
    \includegraphics[width=1.05\linewidth]{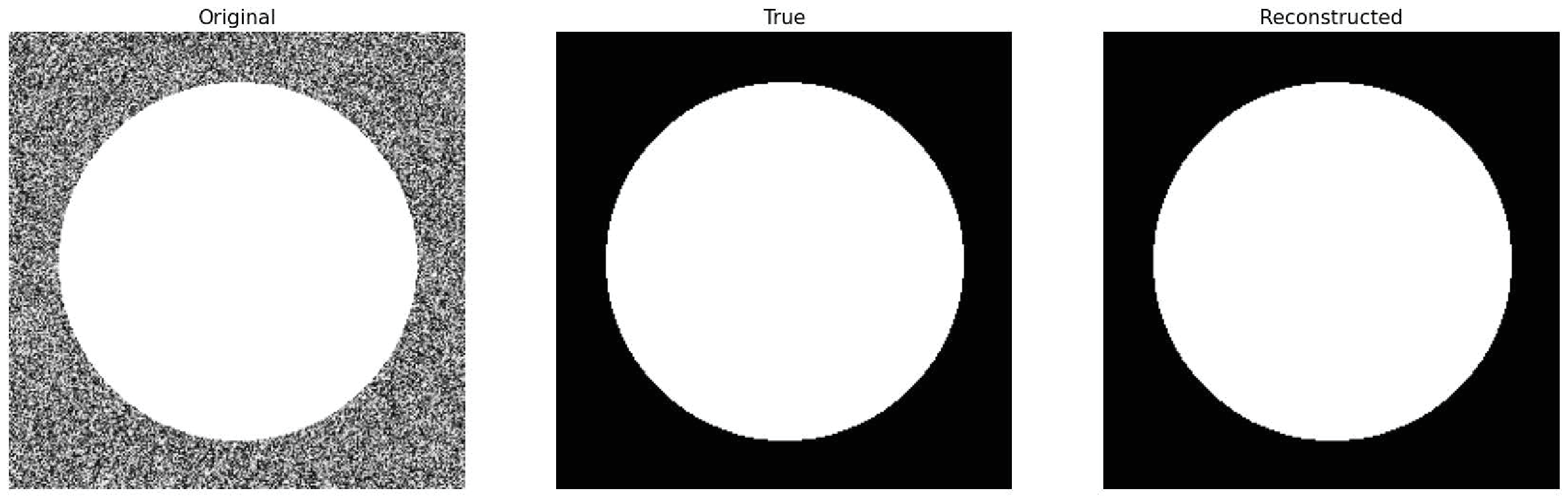} 
    \caption*{\centering (b)}
    \caption{ Images used to test different diffusion functions. The first column displays the original images, the second column presents the expected segmentation mask, and the third column shows the resulting binary mask. Each picture consists of $(256,256)$ pixels. For the optimization procedure we set $T=200$ and $\Delta t = 0.1$. We define the number of iterations at 50. Row (a) shows the image with a square on a blurry background, while row (b) displays a similar image, but with a circle. 
    Only one resulting binary mask was reported for each of the images because all the tests described in this section obtain the same segmentation mask.}
    \label{geo_figures}
\end{figure}

In this section we study the impact of choosing different diffusion functions $D(c)$ in images consisting of a blurry background and a geometric shape in the center, as shown in Figure~\ref{geo_figures}. The objective is to detect the shape of the Geometric Figure and to compare how the choice of different diffusion functions affect the value of the model parameters $\Delta_{1},\Delta_{2}$ and $\sigma^{2}$ where the optimization process is identical to the one introduce in Section \ref{Par.Opt}. To this end, we chose the following diffusion functions:

\begin{equation}
\label{eq:Di}
\begin{aligned}
   &D_{1}(c) = c (1 - c)  &D_{2}(c)= 4 c^{2} (1-c)^{2} \\
    &D_{3}(c) = \begin{cases}
        \frac{c}{2} \hspace{1.8cm}\text{if c} \leq 0.5 \\
        \frac{c}{2} ( 1 - c) \hspace{0.7cm}\text{if c} > 0.5
    \end{cases}  
   &D_{4}(c)= 64 c^{4} (1-c)^{4}. 
\end{aligned}
\end{equation}

\begin{figure}
    \centering
    \includegraphics[scale  = 0.5]{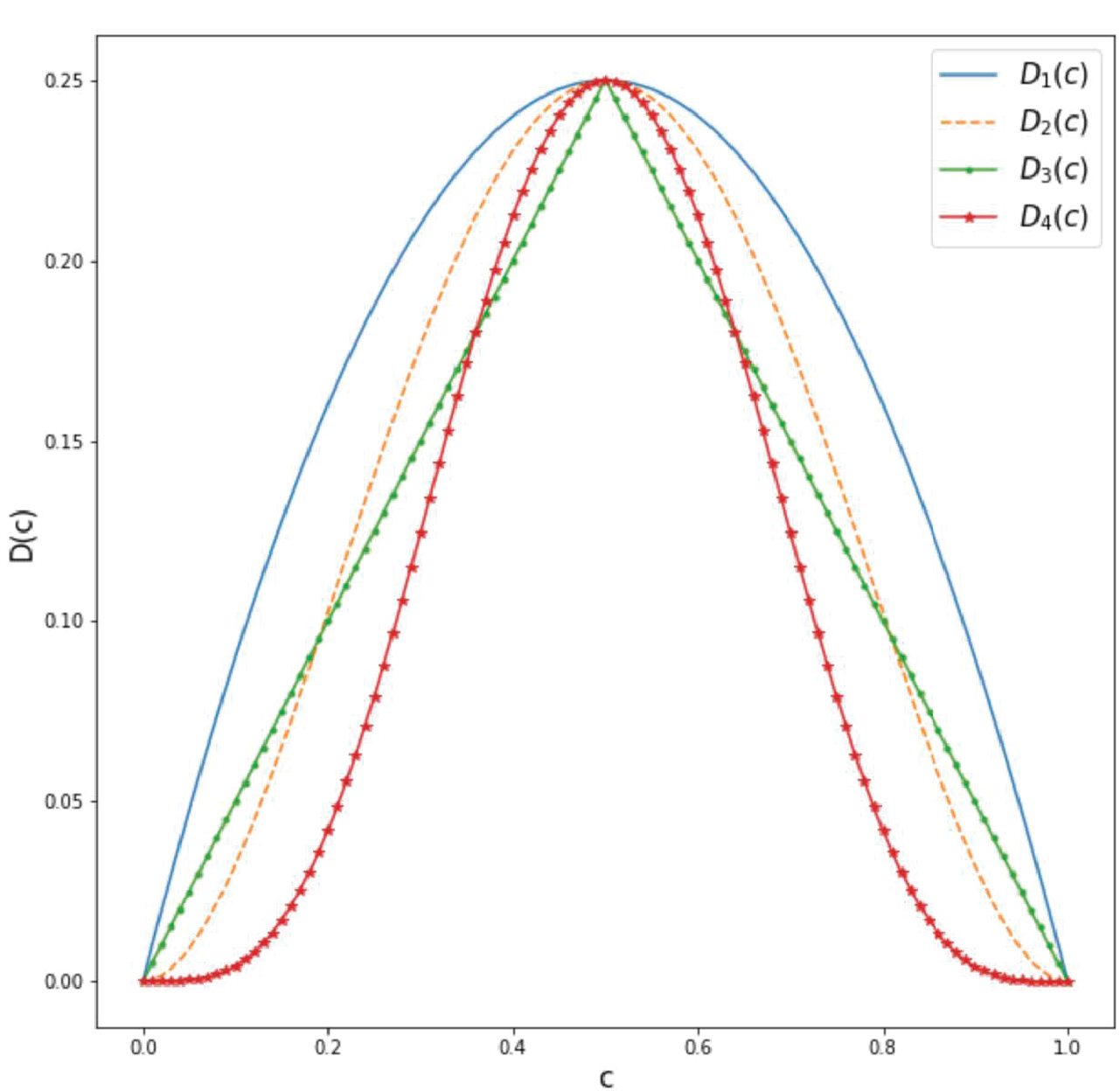}
    \caption{ Diffusion functions defined in \eqref{eq:Di} to assess the variability related to a given feature's level.  }
    \label{fig:dfuncs}
\end{figure}

We point the reader to Figure \ref{fig:dfuncs} for a summary of the various introduced diffusion functions in \eqref{eq:Di}.   

For both the square and circle images, the Surface Dice Coefficient was used to optimize the parameters with a tolerance equal to the length of 1 pixel. Both images have a shape of $(256,256)$ pixels. The final time was set to $T=200$ with $\Delta t = 0.1$. The resulting binary mask was the same for all choices of diffusion functions, obtaining the same loss function value. The results are shown in Figure \ref{geo_figures}. In the case of the square Figure \ref{geo_figures} (a) we can see from Table \ref{square/circle} that for $D_{1}(c)$ and $D_{3}(c)$ the values of $\Delta_{1}$ do not differ greatly for this two diffusion functions. In the case of $\Delta_{2}$ we obtain a slightly smaller value for $D_{1}(c)$ compared to the one obtained for $D_{3}(c)$ and a larger value of the parameter $\sigma^{2}>0$ for $D_{3}(c)$ compared to the one obtained for $D_{1}(c)$.
If we look at Figure~\ref{fig:dfuncs} we notice that $D_{1}(c) \ge D_{3}(c)$. Therefore, a larger value of the diffusion functions is balanced by a smaller value of $\sigma^{2}$ to obtain a similar diffusion effect. This holds also for $D_{1}(c)$ and $D_{3}(c)$ for the circle Figure~\ref{geo_figures} (b). Furthermore, comparing $D_{2}(c)$ and $D_{4}(c)$ for the square image we can see that the resulting parameters are smaller for $D_{2}(c)$ in contrast to the one obtained with $D_{4}(c)$. This is consistent because, again, we can see from Figure \ref{fig:dfuncs} that $D_{2}(c) \ge D_{4}(c)$. If we now compare $D_{2}(c)$ and $D_{4}(c)$ for the circle image we can see that the value of $\sigma^{2}$ is similar in this case. Nevertheless, in this case the difference is given by the values of $\Delta_{1}$ and $\Delta_{2}$, which are both smaller for $D_{2}(c)$.
This indicates that, for different diffusion functions, the optimal parameters adjust to yield similar results. A very straight way is to obtain similar values of $\Delta_{1}$ and $\Delta_{2}$ and a lower value of $\sigma^{2}$ for the diffusion function that has a higher value as in the case of the square image. However, the example of the circle image shows that us that we can also obtain different combinations of parameters so as to counter the effect of a bigger diffusion function.

\begin{table}[H] 
\caption{Parameters obtained for different Diffusion Functions for the Square and Circle Images. The loss metric used to obtain these parameters was the Surface Dice Coefficient with a tolerance equal to the length of 1 pixel. \label{square/circle}}
\newcolumntype{C}{>{\centering\arraybackslash}X}
\begin{tabularx}{\textwidth}{CCCC}
\toprule
\multicolumn{4}{c}{\textbf{Square}} \\ 
\midrule
\ & \textbf{$\Delta_{1}$} & \textbf{$\Delta_{2}$}	& \textbf{$\sigma^{2}$}\\ \\
$D_{1}(c)$  &  0.884	&   0.310	 &   0.889\\
$D_{2}(c)$  &  0.351    &   0.054    &   0.047\\
$D_{3}(c)$  &  0.817    &   0.407    &   1.341\\ 
$D_{4}(c)$  &  0.442	&   0.081	 &   0.624\\ 
\midrule
\multicolumn{4}{c}{\textbf{Circle}} \\
\midrule
\ & \textbf{$\Delta_{1}$} & \textbf{$\Delta_{2}$}	& \textbf{$\sigma^{2}$}\\ \\
$D_{1}(c)$  &  0.435   &   0.341   &   1.829\\
$D_{2}(c)$  &  0.013   &   0.160   &   2.717\\
$D_{3}(c)$  &  0.408   &   0.268   &   2.693\\ 
$D_{4}(c)$  &  0.154   &   0.228   &   2.572\\ 
\bottomrule
\end{tabularx}
\end{table} 

From Table \ref{square/loss} we can see the parameters obtained by minimizing three different optimization metrics using as a diffusion function $D_{1}(c)$ for  the square image. For all the cases the resulting Surface Dice was equal to one indicating a perfect overlap between the computed and the ground truth segmentation masks. 
The resulting binary mask obtained were the same for the three examples and are equivalent to the ones shown in Figure \ref{geo_figures}. 
For the Volumetric and Surface Dice Coefficient we can see that the parameters obtained where identical. Nevertheless, for the Jaccard Index, the resulting parameters differed, being smaller in this case.  The loss is null in both cases, coherently with the relationship \eqref{dice jac rel}.

\begin{table}[H]
\caption{Parameters obtained for the Square Image by minimizing the Jaccard Index and the Volumetric and Surface Dice Coefficient. For the Surface Dice Coefficient the tolerance was set to the length of 1 pixel. The loss obtained was zero for the three cases. \label{square/loss}}
\newcolumntype{C}{>{\centering\arraybackslash}X}
\begin{tabularx}{\textwidth}{CCCC}
\toprule
\multicolumn{4}{c}{\textbf{Square}} \\ 
\midrule
\ & \textbf{$\Delta_{1}$} & \textbf{$\Delta_{2}$}	& \textbf{$\sigma^{2}$}\\ \\
Vol. Dice  &   0.884	&   0.310	 &   0.889\\
Surf. Dice  &   0.884    &   0.310    &   0.889\\ 
JAC        &   0.442 	&   0.081	 &   0.624\\
\bottomrule
\end{tabularx}
\end{table}

\subsection{Determining the final time}

\begin{figure}
    \centering
    \includegraphics[scale=0.7]{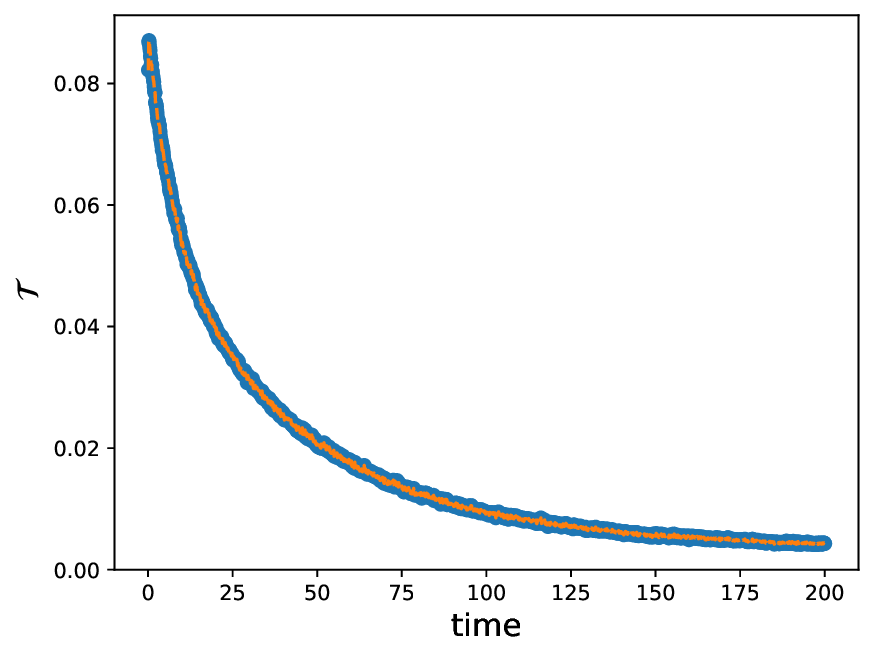}
    \caption{Evolution of $\mathcal{T}$ where the kinetic density is the one considered in  Figure~\ref{geo_figures}(a). The image consists of $(256,256)$ pixels. We can observe how $\mathcal{T}$ decreases until condition $\mathcal{T}<\delta$ is reached with $\delta = 0.005$.}
    \label{fig:steady-state}
\end{figure}

In this section we specify the criterion that we implemented to determine the final time  $T>0$. As defined in Section \ref{sect:DSMC}, we approximate the solution of  \eqref{FP2} through a DSMC approach, even though we have no analytical insight on the form of the steady state.
The objective is to find the values of the final time $T>0$ such that a numerical steady state can be defined. We stress that the time taken to reach the equilibrium state for different initial conditions is not the same so we need to determine the time parameters for all the images we want to analyze. To this end, if $f^n(\mathbf{x},c)$ is the approximation of the density at time $t^{n}= n\Delta t$, we define

\begin{equation}
\mathcal T = \int_{\mathbb{R}^2 \times [0,1]} |f^{n+1}(\mathbf{x},c) - f^n(\mathbf{x},c)|d\mathbf{x}dc,
\end{equation}

which represents an index of variation between two successive time steps of the reconstructed kinetic density. 
As the solution evolves, this quantity decreases and tends to zero as the equilibrium state is reached, as illustrated in Figure \ref{fig:steady-state} for the case of the Square Image with a blurry background. Hence, we may introduce a breaking criterion based on the condition $\mathcal{T} < \delta$, for some $\delta > 0$. When this condition is satisfied, the reconstructed density is considered an approximation of the steady state.

The same procedure was done for all images presented in this work so as to fulfill with the condition presented in this section.

\subsection{Optimization Metrics for Tumor Image Analysis}

\begin{figure}
    \centering
    \includegraphics[width=0.6\linewidth,height=7cm]{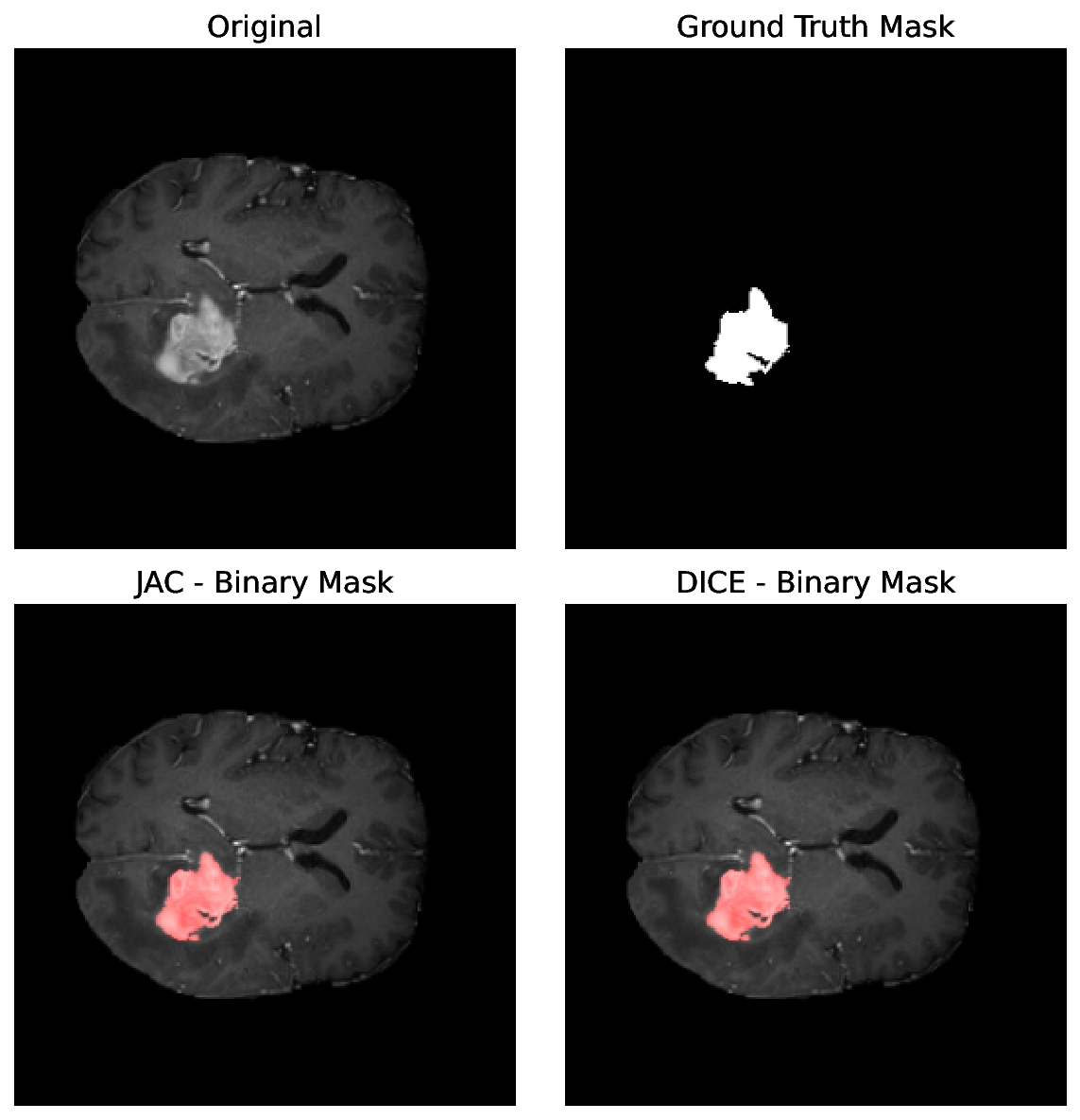}
    \caption*{(a)}
    \vspace{0.1cm}
    \centering
    \includegraphics[width=0.6\linewidth,height=7cm]{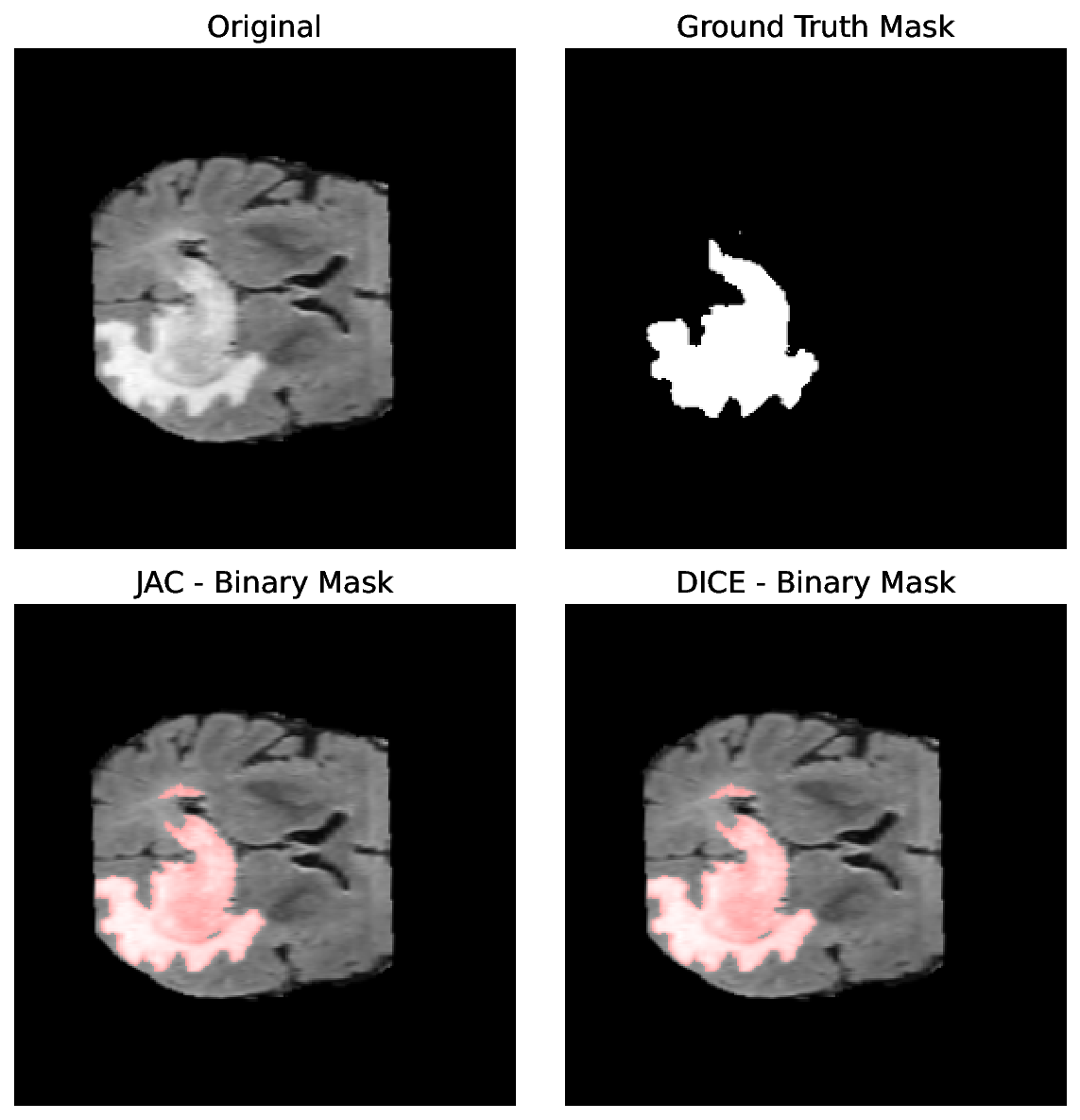}
    \caption*{(b)}
    \caption{Segmentation Masks obtained by minimizing the Jaccard Index and the Volumetric Dice Coefficient. (a) shows the results for the Core Tumor and (b) shows the results for the Whole Tumor. Both images consist of $240\times240$ pixels. For the optimization procedure we set $T=100$ and $\Delta t = 0.01$. In both cases we considered 300 iterations of the optimization algorithm. In both cases the loss reported by the Jaccard Index was smaller compared to the one obtained with the Volumetric Dice Coefficient. Furthermore, it can be noticed that the losses reported satisfy relation \ref{dice jac rel} as expected. From the values of the parameters we can observe that a bigger value of the diffusion is countered by a smaller value of $\Delta_{2}$.}
    \label{fig:jac_dice}
\end{figure}

In this section we study the impact that the different optimization metrics have on the resulting binary mask for the Core and Whole Tumor. We also analyze the parameters obtained for the different optimization metrics. Both brain tumor images consist of $N = (240,240)$ pixels. For the optimization procedure we determine $T = 100$ and $\Delta t = 0.01$. For each segmentation mask generated we evaluated $300$ different combinations of parameters. Figure \ref{fig:jac_dice} show the Segmentation Masks obtained for both the Whole and Core Tumor by optimizing the Jaccard Index and the Volumetric Dice Coefficient. In Table \ref{jac vs dice} the resulting parameters and the loss obtained for both optimization metrics are presented, in this case the loss is equal to 1 for a perfect overlap and 0 if the images are totally disjoint. First, we can observe that the loss obtained with both metrics satisfy  \eqref{dice jac rel} as expected. It can be noticed that for both segmentation masks the loss obtained is greater for the Volumetric Dice Coefficient. Furthermore, the parameter $\Delta_{1}$ obtained with both optimization metrics is similar for both the Core and Whole Tumor. Nevertheless, we can see that for the Whole Tumor the $\Delta_{2}$ parameter obtained with the Jaccard Index is bigger than the one obtained with the Volumetric Dice Coefficient. For the case of the Core Tumor instead the $\Delta_{2}$ parameter is bigger for the Volumetric Dice Coefficient. If we compare this to the values obtained for $\sigma^{2}$ in both cases for both metrics we can see that a bigger diffusion value is countered by a smaller value of $\Delta_{2}$ so as to obtain similar Segmentation Masks as seen from Figure \ref{fig:jac_dice}.

\begin{figure}
    \centering
    \includegraphics[width=0.6\linewidth,height=7cm]{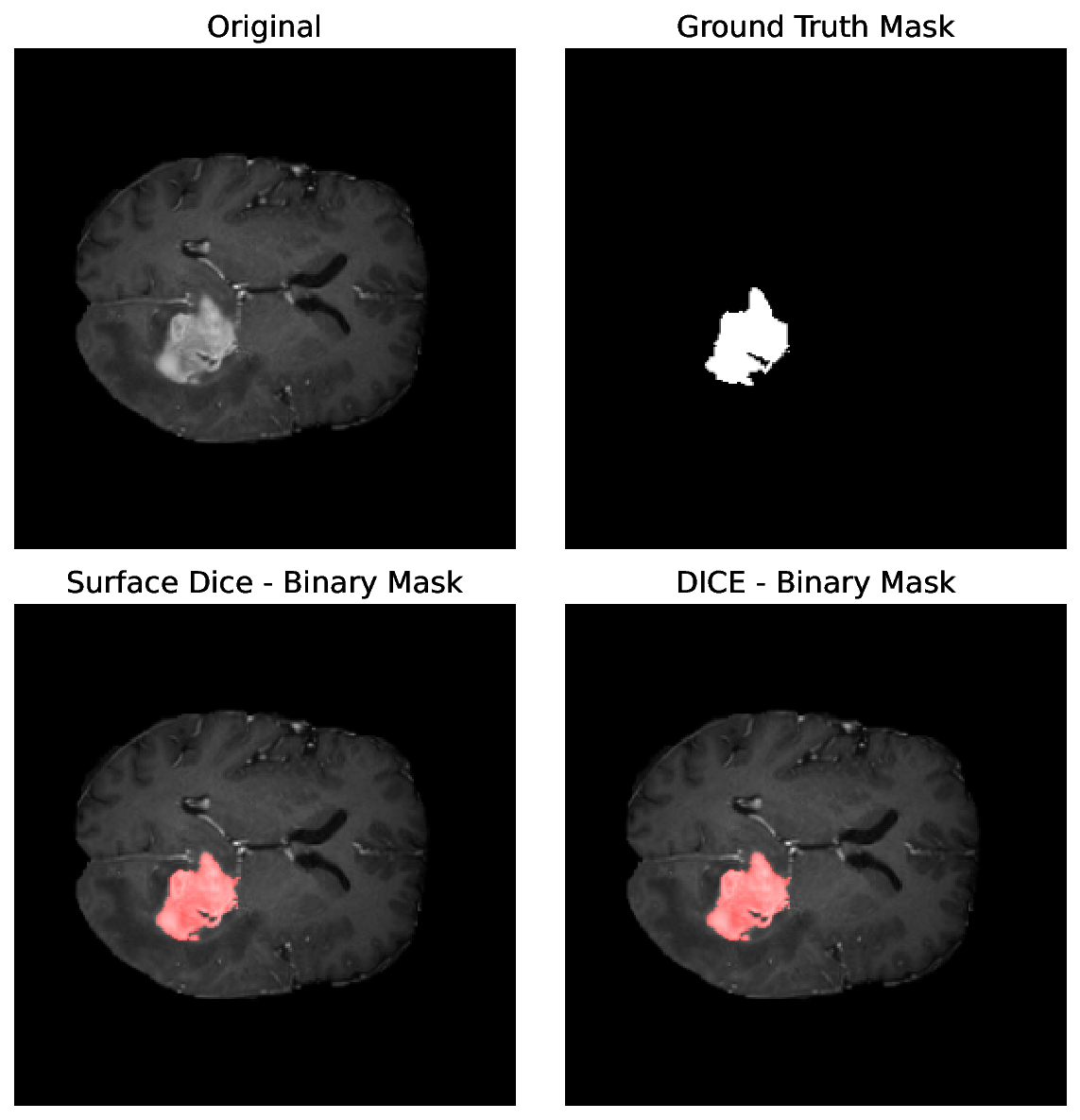}
    \caption*{(a)}
    \vspace{0.1cm}
    \centering
    \includegraphics[width=0.6\linewidth,height=7cm]{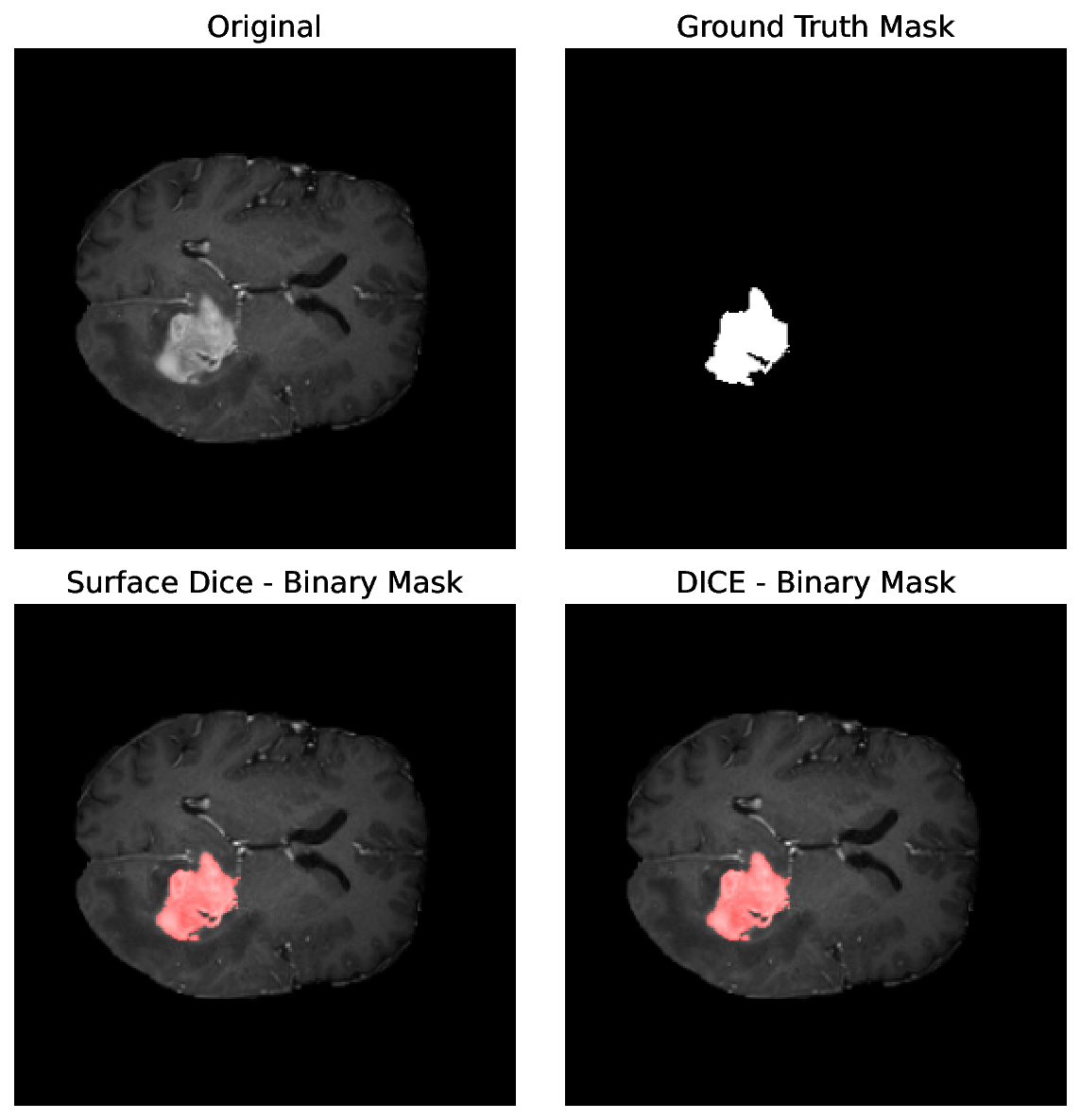}
    \caption*{(b)}
    \caption{Segmentation Masks obtained by minimizing the Surface and Volumetric Dice Coefficient. (a) shows the results for the Core Tumor and (b) shows the results for the Whole Tumor. Both images consist of $240\times 240$ pixels. For the optimization procedure we set $T=100$ and $\Delta t = 0.01$. In both cases we considered 300 iterations of the optimization algorithm. For the Surface Dice Coefficient we set the tolerance $\tau$ equal to the length of 1 pixel. Given that both the Volumetric Dice Coefficient and the Jaccard Index are a measure of the superposition between the two surfaces and do not account for the proximity between the two surfaces at every given point the Surface Dice Coefficent represents a more suitable metric when comparing two different surfaces.}
    \label{fig:surface}
\end{figure}

For the Surface Dice Coefficient, the tolerance $\tau$ was set to the length of 1 pixel, both when used as the optimization loss and when used as the evaluation metric. In Figure \ref{fig:surface} shows the resulting binary mask obtained with the Surface Dice Coefficient and the Volumetric Dice Coefficient for the Core and Whole Tumor. In the case of the Whole Tumor the loss obtained with Surface Dice Coefficient is smaller than the one obtained with the Jaccard Index and the Volumetric Dice Coefficient. For the Core Tumor the loss obtained with the Surface Dice Coefficient is similar to the one reported by the Jaccard Index and both are smaller than the obtained with the Volumetric Dice Coefficient. For the Whole Tumor  we can see that the resulting parameters are similar for all the optimization metrics. Nevertheless, for the Core Tumor we can notice that the parameters obtained with the Surface Dice Coefficient differ compared to the ones obtained with the Jaccard Index and the Volumetric Dice Coefficient. In particular, we obtained a smaller value for $\sigma^{2}$ and slightly bigger value for $\Delta_{1}$. This indicates that a smaller value for the diffusion of the particles is compensated by allowing the particles to aggregate with others that are slightly more separated than in the case of the Volumetric Dice and Jaccard Index. Given that both the Volumetric Dice Coefficient  and the Jaccard Index are a measure of the superposition between two volumes (in this case two surfaces) they do not represent the proximity between two surfaces making the Surface Dice Coefficient more suitable to use as a loss metric when comparing two different surfaces.

\begin{table}
\caption{Parameters obtained for the Whole and Core Tumor using the Volumetric Dice Coefficient, Jaccard Index and the Surface Dice Coefficient. The loss reported is 1 for perfect overlap and 0 for complete deviation. \label{jac vs dice}}
\newcolumntype{C}{>{\centering\arraybackslash}X}
\begin{tabularx}{\textwidth}{CCCCC}
\toprule
\multicolumn{5}{c}{\textbf{Whole Tumor}} \\ 
\midrule
Opt. Function & \textbf{$\Delta_{1}$} & \textbf{$\Delta_{2}$}	& \textbf{$\sigma^{2}$} & Loss\\ \\
Vol. Dice     &     0.4972	&   0.0888	 &   2.6867   &   0.9292\\
JAC      &     0.5075   &   0.1187   &   2.3631   &   0.8672\\
Surf. Dice &     0.6383   &   0.0579   &   2.6504   &   0.7447\\ 
\midrule
\multicolumn{5}{c}{\textbf{Core Tumor}} \\
\midrule
Opt. Function & \textbf{$\Delta_{1}$} & \textbf{$\Delta_{2}$}	& \textbf{$\sigma^{2}$} & Loss\\ \\
Vol. Dice     &     0.3795	&   0.1254	 &   2.1808   &   0.9360\\
JAC      &     0.3823   &   0.1004   &   2.7001   &   0.8796 \\ 
Surf. Dice &     0.6841   &   0.0760   &   1.4155   &   0.8727 \\ 
\bottomrule
\end{tabularx}
\end{table}

\begin{figure}
    \centering
    \includegraphics[scale = 0.22]{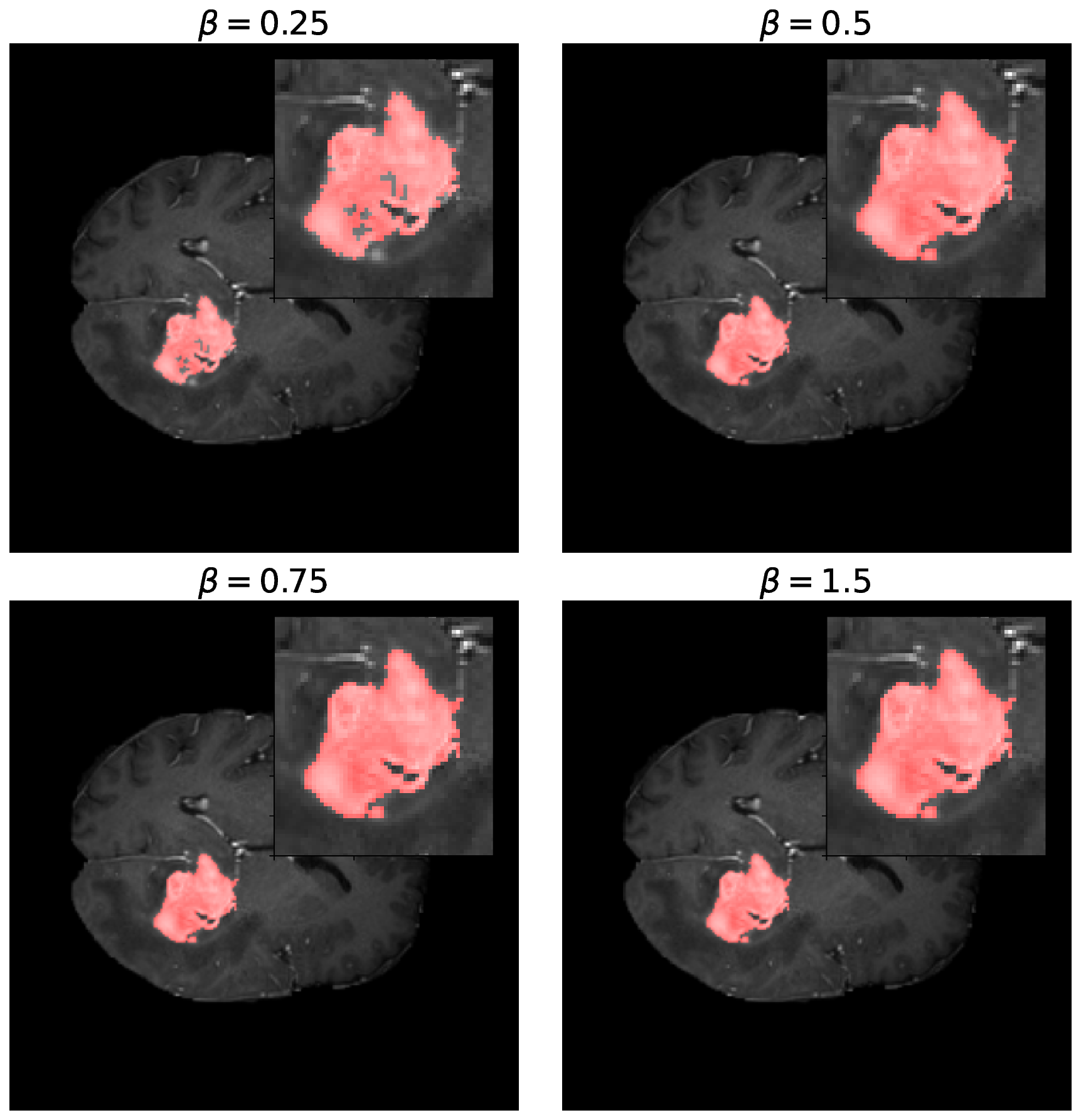}
    \caption*{\centering (a)}
    \vspace{0.3cm}
    \centering
    \includegraphics[scale = 0.22]{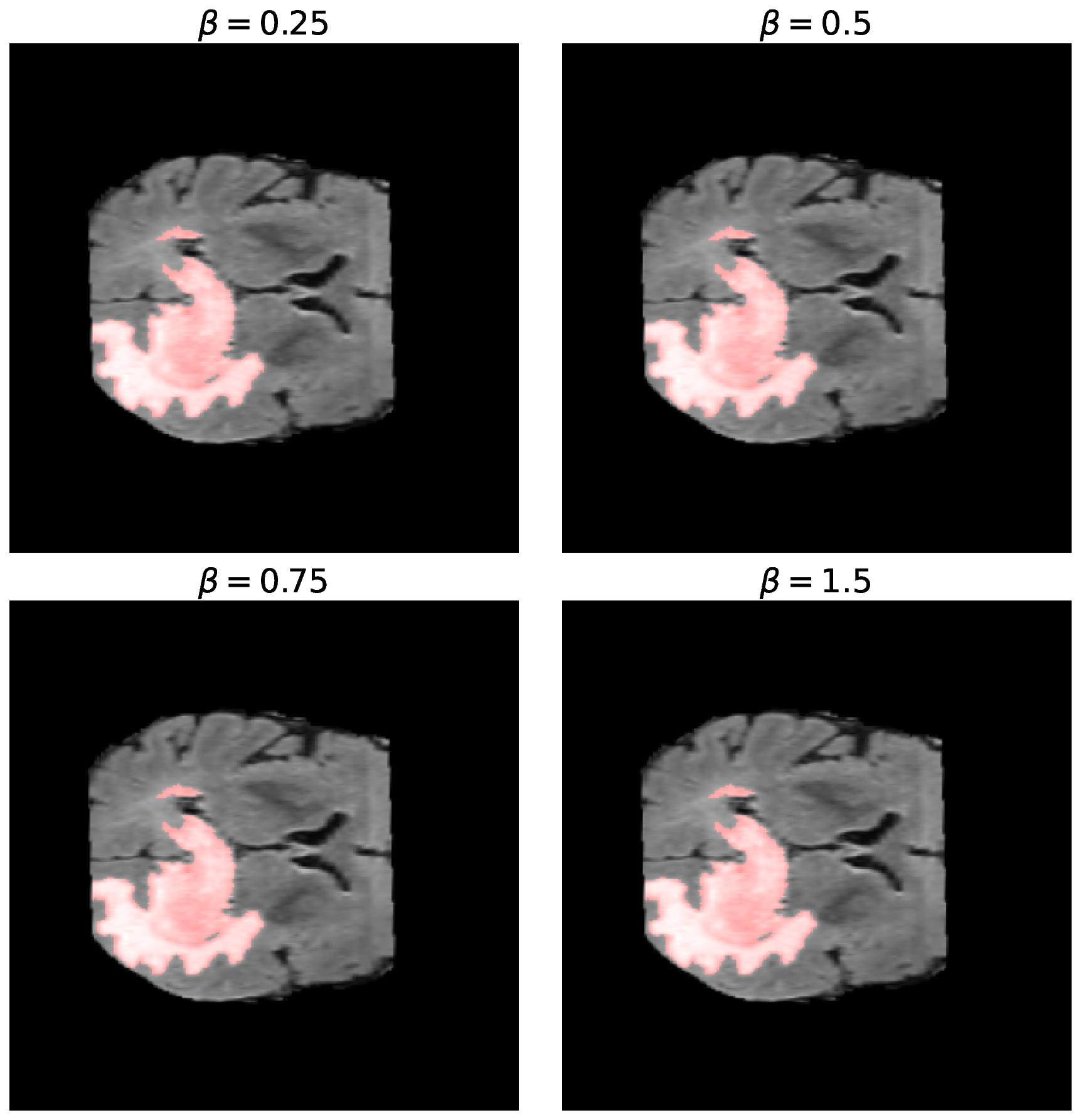}
    \caption*{\centering  (b)}
    \caption{Segmentation masks obtained for the $F_{\beta}-$Loss metric. (a) Shows the segmentation masks obtained for $\beta = 0.25,0.5,0.75,1.5$ for the Core tumor and (b) Shows the segmentation masks obtained using the same values of $\beta$ for the Whole Tumor. Both images consist of $240\times 240$ pixels. For the optimization procedure we set $T=300$ and $\Delta t = 0.01$. In both cases we considered 300 iterations of the optimization algorithm. In (a) we can observe that for $\beta=0.25$ the resulting segmentation masks display areas of misclassified pixels while for bigger values of $\beta$ the resulting segmentation mask does not differ. In (b) no zoomed area is shown as the segmentation masks display no visible differences for the different values of $\beta$. This can be seen also from Table~\ref{F measure} by observing the number of False Positives (FP), False Negatives (FN) and True Positives (TP) obtained for both images.}
    \label{fig:betabm}
\end{figure}

\begin{figure}
    \centering
    \includegraphics[scale = 0.4]{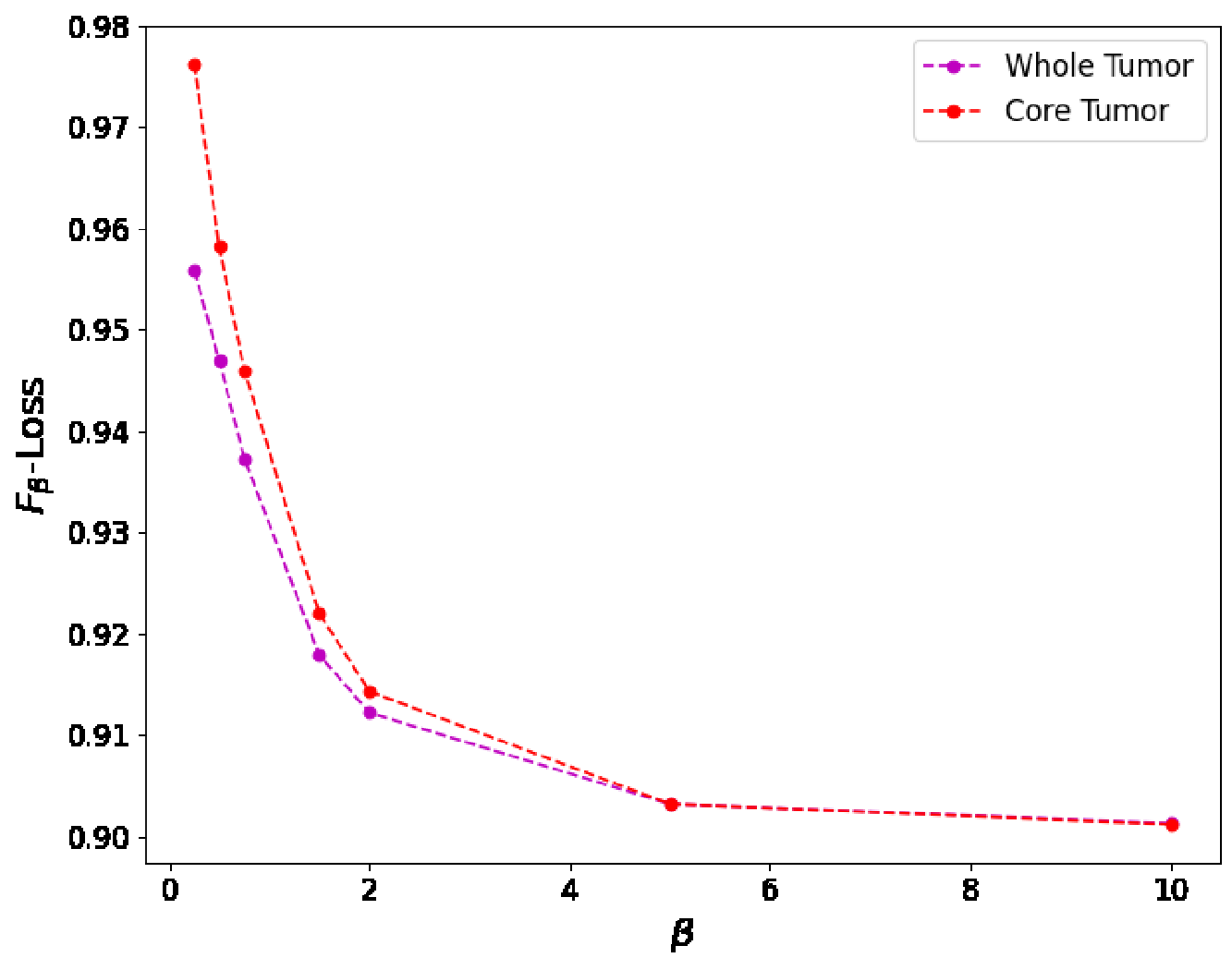}
    \caption{Relationship between the F$_\beta-$ loss value and the $\beta$ value for both the Core and Whole Tumor Images. As $\beta$ increases, the F$_\beta-$ loss decreases showing that for lower values of $\beta$ we should obtain a more precise segmentation mask as the loss indicated in this Figure is 1 for perfect overlap. Nevertheless, the resulting binary mask is less accurate for lower values of $\beta$ showing that this is not an appropriate metric for optimizing the consensus-based model.}
    \label{fig:betaloss}
\end{figure}

For the $F_{\beta}-$measure we can see in Figure~\ref{fig:betabm} the binary masks obtained for different values of $\beta$ for the Core and Whole Tumor. For the case of the Core Tumor we can observe that for $\beta = 0.25$ we obtain areas of misclassified pixels in the tumor region. This can also be seen from Table \ref{F measure} where the number of False Negatives is bigger and the number of False Positives is smaller compared to the results obtained for bigger values of $\beta$. If we recall \eqref{F_measure_rel} we can see that for low values of $\beta$ the False Negatives are multiplied by a factor of $\beta^{2}$ thus having a smaller weight compared to the False Positives. As we increase the value of $\beta$ we can notice from both Table \ref{F measure} and Figure~\ref{fig:betabm} that modifying the value of $\beta$ has no impact on the resulting binary mask. This also holds true for the Whole Tumor as no difference can be noticed in the results obtained for different values of $\beta$. Finally in Figure~\ref{fig:betaloss} we see the loss reported for different values of $\beta$ where  the loss equal to $1$ represents a perfect overlap. First, it can be noticed that we obtain the higher value of the loss for $\beta=0.25$, meaning that this should be the most accurate result, which is anyway balanced by the fact that we obtain the larger number of False Negatives. Again, we saw that this can be obtained from \eqref{F_measure_rel}, where low values of $\beta$ reduce the impact of a large number of False Negatives on the resulting loss. Secondly, we observe that the loss decreases for larger values of $\beta$. This behavior arises because the loss is inversely proportional to $\beta$, while the resulting segmentation masks remain unchanged, as shown in Table \ref{F measure}. This shows that the $F_{\beta}$-measure may not be a reliable metric for these types of segmentation masks and this segmentation method, and that modifying the value of $\beta$ provides no advantage.

\begin{table}[H] 
\caption{ Parameters obtained for the $F_{\beta}$-measure for different values of $\beta$. The loss reported is 1 for perfect overlap and 0 for complete deviation. The number of False Positives (FP), False Negatives (FN) and True Positives (TP) are presented for the resulting segmentation mask for each value of $\beta$. \label{F measure}}
\newcolumntype{A}{>{\centering\arraybackslash}X}
\resizebox{\textwidth}{!}{  
\begin{tabularx}{1.2\textwidth}{AAAAAAAA}
\toprule
\multicolumn{8}{c}{\textbf{Whole Tumor}} \\ 
\midrule
\ & \textbf{$\Delta_{1}$} & \textbf{$\Delta_{2}$}	& \textbf{$\sigma^{2}$} & FP & FN & TP & Loss\\ \\
$\beta = 0.25$     &     0.6873	  &   0.1707   &   2.2395   &  134 & 347 & 3170 & 0.9559\\
$\beta = 0.5$      &     0.3351   &   0.1080   &   2.7051   & 134 & 350 & 3167 & 0.9470\\
$\beta = 0.75$     &     0.5939   &   0.2304   &   2.6718   & 134 & 350 & 3167 & 0.9373\\
$\beta = 1.5$     &     0.5316   &   0.1092   &   2.7105   & 136 & 349 & 3168 & 0.9179\\
$\beta = 5.0$     &     0.5662   &   0.1225   &   2.7043   &  136 & 349 & 3168 & 0.9032\\
$\beta = 10.0$     &     0.6061   &   0.2835   &   2.1243   &  136 & 349 & 3168 & 0.9013\\
\midrule
\multicolumn{8}{c}{\textbf{Core Tumor}} \\
\midrule
\ & \textbf{$\Delta_{1}$} & \textbf{$\Delta_{2}$}	& \textbf{$\sigma^{2}$} & FP & FN & TP & Loss\\ \\
$\beta = 0.25$     &     0.6575	  &   0.2725   &   0.0257   &   9 & 206 & 849 & 0.9763\\
$\beta = 0.5$      &     0.3989   &   0.0637   &   1.8094   &   25 & 107 & 948 & 0.9582\\
$\beta = 0.75$     &     0.4073   &   0.0942   &   1.6972   &    25 & 105 & 950 & 0.9460\\
$\beta = 1.5$      &     0.5444   &   0.2077   &   2.3545   &   25 & 105 & 950 & 0.9220\\
$\beta = 5.0$      &     0.5587   &   0.1742   &   2.6864   &  25 & 105 & 950 & 0.9032\\
$\beta = 10.0$     &     0.6137   &   0.2425   &   1.9757   &  25 & 105 & 950 & 0.9012\\
\bottomrule
\end{tabularx}
}
\end{table}

\section*{Conclusions}

In this paper we presented a consensus-based kinetic method and show how can this model can be applied for the problem of image segmentation. The pixel in a 2D image is interpreted as a particle that interacts with the rest through a consensus-type process, which allows us to identify different clusters and generate an image segmentation. We developed a procedure that allows us to approximate the Ground Truth Segmentation Mask of different Brain Tumor Images. Furthermore, we presented and evaluated different optimization metrics and study the impact on the results obtained. In particular we found that the Jaccard Index and the Volumetric and Surface Dice Coefficient are appropriate metric to optimize our model. Nevertheless, given that the Surface Dice Coefficient is measure of discrepancy between the boundaries of two surfaces it is a better representation compared to the Jaccard Index and the Volumetric Dice Coefficient as they account only for absolute differences and do not attain to point wise differences. Furthermore, we assessed the use of the $F_{\beta}$-Loss as a potential optimization metric. We found that both the loss values and the corresponding results were difficult to interpret, as low loss values often corresponded to low accuracy, making this metric challenging to apply effectively for optimization in this context. Future researches will focus on the case of multidimensional features and potential training methods for the introduced model.

\vspace{6pt} 



\section*{Acknowledgments}
M.Z. is member of GNFM (Gruppo Nazionale di Fisica Matematica) of INdAM, Italy and acknowledges support of PRIN2022PNRR project No.P2022Z7ZAJ, European Union - NextGeneration EU. M.Z. acknowledges partial  support by ICSC – Centro Nazionale di Ricerca in High Performance Computing, Big Data and Quantum Computing, funded by
European Union – NextGenerationEU


\end{document}